\documentclass[12pt]{article}
\usepackage{comment,array}
\usepackage{booktabs,multirow,xcolor,subcaption}
\usepackage{caption,setspace}
\usepackage{threeparttable}
\usepackage[colorlinks,bookmarksopen,bookmarksnumbered,citecolor=blue,urlcolor=blue,linkcolor=blue]{hyperref}
\usepackage{float}
\usepackage{amsmath}
\usepackage{pifont}

\usepackage{afterpage}    
\usepackage{pdflscape} 

\newcommand{\argmin}{\ensuremath{\operatornamewithlimits{argmin}}}

\newcommand{\rank}{\mathrm{rank}}

\newcommand{\domega}{{\rm d}\omega}

\def\spacingset#1{\renewcommand{\baselinestretch}%
	{#1}\small\normalsize} \spacingset{1}

\renewcommand{\arraystretch}{1.3}  

\usepackage{xcolor}
\usepackage[draft,inline,nomargin,index]{fixme}
\fxsetup{theme=color,mode=multiuser}
\FXRegisterAuthor{yf}{ayf}{\color{magenta}}

\usepackage{amsthm,amsmath,enumerate,amsbsy,amsfonts,amssymb,mathabx,amscd,graphicx,algorithm, algpseudocode, indentfirst}
\usepackage{natbib}
\usepackage{geometry}
\usepackage{authblk, caption, subcaption}
\usepackage{mathrsfs}
\usepackage{epsfig}
\newtheorem{definition}{Definition}
\newtheorem{theorem}{Theorem}

\newtheorem{remark}{Remark}

\newtheorem{condition}{Condition}
\newtheorem{assumption}{Assumption}

\usepackage{multirow}

\newcommand{\bzero}{\boldsymbol 0}

\newcommand{\bOmega}{\boldsymbol \Omega}
\newcommand{\bSigma}{\boldsymbol \Sigma}

\newcommand{\bTheta}{\boldsymbol \Theta}

\newcommand{\bDelta}{\boldsymbol \Delta}

\newcommand{\bmu}{\boldsymbol \mu}

\newcommand{\bfeta}{\boldsymbol \eta}
\newcommand{\bvar}{\boldsymbol \varepsilon}

\newcommand{\bvarepsilon}{\boldsymbol \varepsilon}

\newcommand{\bpi}{\boldsymbol \pi}
\newcommand{\bPi}{\boldsymbol \Pi}

\newcommand{\bd}{{\mathbf d}}
\newcommand{\be}{{\mathbf e}}
\newcommand{\bbf}{{\mathbf f}}
\newcommand{\bx}{{\mathbf x}}
\newcommand{\by}{{\mathbf y}}
\newcommand{\bu}{{\mathbf u}}
\newcommand{\bv}{{\mathbf v}}

\newcommand{\bw}{{\mathbf w}}

\newcommand{\bz}{{\mathbf z}}

\newcommand{\bD}{{\bf D}}
\newcommand{\bA}{{\bf A}}
\newcommand{\bB}{{\bf B}}
\newcommand{\bC}{{\bf C}}

\newcommand{\bI}{{\bf I}}
\newcommand{\bL}{{\bf L}}

\newcommand{\bP}{{\bf P}}

\newcommand{\bX}{{\bf X}}

\newcommand{\bU}{{\bf U}}
\newcommand{\bV}{{\bf V}}

\newcommand{\bH}{{\bf H}}

\newcommand{\cC}{{\cal C}}
\newcommand{\cD}{{\cal D}}
\newcommand{\cE}{{\cal E}}
\newcommand{\cI}{{\cal I}}
\newcommand{\cL}{{\cal L}}
\newcommand{\cM}{{\cal M}}
\newcommand{\cG}{{\cal G}}

\newcommand{\cT}{{\cal T}}
\newcommand{\cU}{{\cal U}}

\newcommand{\cS}{{\cal S}}

\newcommand{\cP}{{\cal P}}
\newcommand{\cN}{{\cal N}}
\newcommand{\cH}{{\cal H}}
\newcommand{\cQ}{{\cal Q}}
\newcommand{\cX}{{\cal X}}
\newcommand{\cY}{{\cal Y}}
\newcommand{\cF}{{\cal F}}

\newcommand{\eZ}{\mathbb{Z}}
\newcommand{\eR}{\mathbb{R}}

\newcommand{\eE}{\mathbb{E}}

\newcommand{\eP}{\mathbb{P}}
\newcommand{\eC}{\mathbb{C}}

\newcommand{\ti}{\text{i}}

\newcommand{\svd}{\text{svd}}
\newcommand{\reg}{\text{reg}}

\newcommand{\tpa}{\text{pa}}
\newcommand{\tch}{\text{ch}}
\newcommand{\tne}{\text{ne}}
\newcommand{\tr}{\mbox{tr}}
\newcommand{\diag}{\mbox{diag}}
\newcommand{\vvec}{{\rm vec}}

\newcommand{\gsupp}{{\rm gsupp}}
\newcommand{\grank}{{\rm grank}}
\newcommand{\sign}{{\rm sign}}
\newcommand{\col}{{\rm col}}

\def\T{{ \mathrm{\scriptscriptstyle T} }}
\def\H{{ \mathrm{\scriptscriptstyle H} }}
\def\F{{ \mathrm{\scriptstyle F} }}

\def\hard{{ \mathrm{\scriptstyle hard} }}

\makeatletter
\newcommand*{\rom}[1]{\expandafter\@slowromancap\romannumeral #1@}
\makeatother

\makeatletter
\DeclareRobustCommand\widecheck[1]{{\mathpalette\@widecheck{#1}}}
\def\@widecheck#1#2{%
	\setbox\z@\hbox{\m@th$#1#2$}%
	\setbox\tw@\hbox{\m@th$#1%
		\widehat{%
			\vrule\@width\z@\@height\ht\z@
			\vrule\@height\z@\@width\wd\z@}$}%
	\dp\tw@-\ht\z@
	\@tempdima\ht\z@ \advance\@tempdima2\ht\tw@ \divide\@tempdima\thr@@
	\setbox\tw@\hbox{%
		\raise\@tempdima\hbox{\scalebox{1}[-1]{\lower\@tempdima\box
				\tw@}}}%
	{\ooalign{\box\tw@ \cr \box\z@}}}
\makeatother
\RequirePackage[colorlinks,citecolor=blue,urlcolor=blue]{hyperref}

\newcommand{\blind}{1}

\setcitestyle{aysep={,},yysep={;}}

\begin{document}
	\if1\blind
	{
 	{
  \spacingset{1.25}
		\title{\bf \Large
        Time Series Gaussian Chain Graph Models
    \hspace{.2cm}\\
        }
		\author[1]{Qin Fang}
		\author[2]{Xinghao Qiao}
		\author[3]{Zihan Wang}
           \affil[1]{\it University of Sydney Business School, Sydney, Australia}
           \affil[2]{\it Faculty of Business and Economics, The University of Hong Kong, Hong Kong SAR}
           \affil[3]{\it Department of Statistics and Data Science, Tsinghua University, Beijing, China}
		\setcounter{Maxaffil}{0}
		
		\renewcommand\Affilfont{\itshape\small}
		\date{\vspace{-5ex}}
		\maketitle
	} \fi
	\if0\blind
	{  \spacingset{1.7}
		\bigskip
		\bigskip
		\bigskip
		\begin{center}
			{\Large\bf Time Series Gaussian Chain Graph Models 
            }
		\end{center}
		\medskip
	} \fi

\setcounter{Maxaffil}{0}
\renewcommand\Affilfont{\itshape\small}
\spacingset{1.5}
\begin{abstract}

Time series graphical models have recently received considerable attention for characterizing (conditional) dependence structures in multivariate time series. In many applications, the multivariate series exhibit variable-partitioned blockwise dependence, with distinct patterns within and across blocks. In this paper, we introduce a new class of time series Gaussian chain graph models that represent contemporaneous and lagged causal relations via directed edges across blocks, while capturing within-block conditional dependencies through undirected edges. In the frequency domain, this formulation induces a cross-frequency shared group sparse plus group low-rank decomposition of the inverse spectral density matrices, which we exploit to establish identifiability of the time series chain graph structure. Building on this, we then propose a three-stage learning procedure for estimating the undirected and directed edge sets, which involves optimizing a regularized Whittle likelihood with a group lasso penalty to encourage group sparsity and a novel tensor-unfolding nuclear norm penalty to enforce group low-rank structure. We investigate the asymptotic properties of the proposed method, ensuring its consistency for exact recovery of the chain graph structure. The superior empirical performance of the proposed method is demonstrated through both extensive simulation studies and an application to U.S. macroeconomic data that highlights key monetary policy transmission mechanisms.

\end{abstract}

\bigskip \bigskip
\noindent%
\bigskip
\noindent








\newpage
\spacingset{1.7}
\setlength{\abovedisplayskip}{0.2\baselineskip}
\setlength{\belowdisplayskip}{0.2\baselineskip}
\setlength{\abovedisplayshortskip}{0.2\baselineskip}
\setlength{\belowdisplayshortskip}{0.2\baselineskip}

\section{Introduction}
\label{sec:intro}

Graphical modelling for multivariate time series has attracted growing interest for its ability to characterize various (conditional) dependence structures among component series, with applications across scientific and economic domains such as environmental science \citep{dahlhaus2003causality}, functional genomics \citep{Shojaie2012}, neuroscience \citep{Foti2016} and financial economics \citep{Lin2017}.
These data can be represented as a stationary $p$-dimensional time series $\bx_t = (x_{t1}, \dots, x_{tp})^\T,$ observed for $t \in [T] :=  \{1, \dots, T\}.$

Existing literature on time series graphical models can be broadly divided into two categories. The first focuses on undirected graphs, where edges represent the conditional dependence structure among $p$ component series of $\{\bx_t\}_{t \in [T]}.$ In the Gaussian setting, this amounts to identifying the nonzero entries of the inverse spectral density matrices \citep{dahlhaus2000graphical}, leading to a frequency-domain representation of the conditional independence graph (CIG). See \cite{jung2015graphical,tugnait2022sparse} for related CIG learning methods.
The second category considers mixed graphs, where directed edges capture dynamic (lagged) Granger-causal relations \citep{granger1969}, and undirected edges encode contemporaneous conditional dependencies. This gives rise to the Granger causality graph \citep{Eichler2007,eichler2012graphical}, where  each component series $\{x_{tj}\}_{t \in [T]}$ is represented by a single node $j$. This formulation is closely related to vector autoregressive (VAR) models, where directed and undirected edges are respectively encoded by the nonzero entries of the transition coefficient matrices and the precision matrix of Gaussian innovations. Variants of the VAR-based representation and the associated learning methods have been proposed, see, e.g., \cite{Basu2015,Lin2017,barigozzi2019nets, Barigozzi2024}.

Alternatively, $\{\bx_t\}$ may be represented by a time-indexed chain graph \citep{dahlhaus2003causality}, in which each node corresponds to a component series at a specific time point $x_{tj}$ yielding a total of $pT$ nodes.  Undirected edges represent contemporaneous conditional dependencies, thereby inducing a natural block structure in which the $p$ nodes from the same time $t$ form one block (i.e., chain component).
Directed edges are allowed only across blocks in temporal order, depicting dynamic causal relations.
In many applications, however, interest extends beyond this time-indexed chain graph to settings where the $p$ component series themselves can be grouped into meaningful blocks. In financial economics, previous work has shown that changes in policy interest rates
can trigger sizable movements in stock prices  over short horizons 
\citep{BernankeKuttner2005}. 
Meanwhile, the interest-rate variables and the asset-return series display both contemporaneous and dynamic conditional dependencies within their respective blocks \citep{AngPiazzesi2003}. 
A similar dependence structure arises in neuroscience, where empirically identified brain functional networks (e.g., frontoparietal, visual) exhibit strong within-network connectivity and directed interactions across networks \citep{Power2011}.

Under an i.i.d. setting, such variable-partitioned blockwise dependence patterns can be represented by classical chain graphs. Introduced as a generalization of directed acyclic graphs (DAGs) characterizing causal relations and undirected graphs depicting the conditional dependence structure, chain graphs \citep{lauritzen1989graphical} admit both directed and undirected edges in one graph by partitioning variables into chain components (i.e., blocks), where cross-block causal relations are encoded via directed edges and within-block conditional dependencies are captured via undirected edges. 
Notably, \citet{zhao2024identifiability} establish identifiability and consistent estimation for Gaussian chain graphs under the Andersson–Madigan–Perlman (AMP) interpretation \citep{andersson2001alternative} via a linear structural equation model.
However, such a formulation is not directly applicable to time series data, as it ignores the dynamic causal relations and leaves the dynamic conditional dependence structure unaddressed.



Our paper introduces a new class of time series chain graphs to capture blockwise causal relations and conditional dependencies, providing a practically useful and interpretable framework for graphical modelling of multivariate Gaussian time series. To this end, we assume the AMP Markov property \citep{andersson2001alternative} and propose model~\eqref{eq:model} to formulate the chain graph structure.
The causal relations (both contemporaneous and dynamic) and the remaining conditional dependencies are encoded via the zero patterns of the coefficient matrices ($\bA$ and $\bB$) and the inverse spectral density matrices of the Gaussian noise process $\{\be_t\}_{t \in [T]}$, respectively, which in turn determine the directed and undirected edges of the proposed time series Gaussian chain graph. 
Figure~\ref{fig:three_plots} provides a toy example, where nodes with different colors represent different chain components, i.e., $\{1,3,5\}, \{2,4\}, \{6\}, \{7\}$. Specifically, the conditional dependencies within each chain component correspond to an undirected CIG, and the causal relations between chain components follow a DAG. 
To capture the dynamics of time series,  we work in the frequency domain and adopt a new three-way tensor representation that encodes temporal dependence along the frequency mode.  The inverse spectral density matrices then admit a cross-frequency shared group sparse plus group low-rank decomposition, which we leverage to establish a  novel identifiability framework.

\begin{figure}[t]
  \centering
  \newsavebox{\measurebox}
  \sbox{\measurebox}{%
    \begin{minipage}{0.4\linewidth}
 \hspace{1.5em}
        \includegraphics[height=8.6cm]{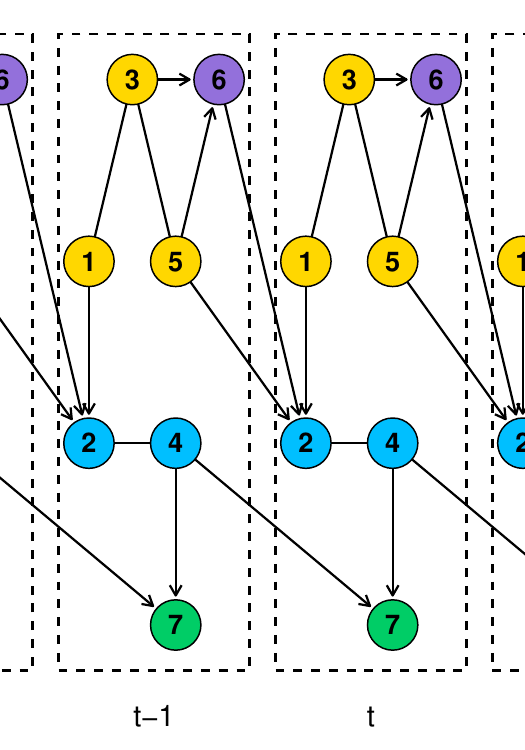}
    \end{minipage}
  }
  
  \usebox{\measurebox}
  \hfill
  \begin{minipage}{0.55\linewidth}
    \hspace{-1.5em}
    \begin{subfigure}{\linewidth}
      \centering
\includegraphics[height=3.8cm]{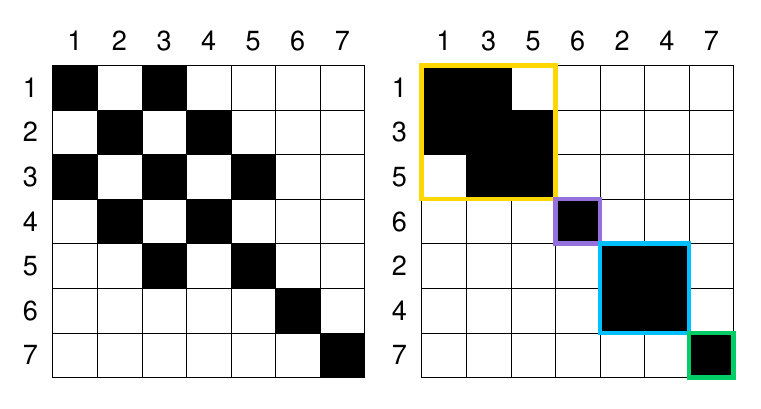}
      \vspace{-0.3cm}
      \caption{$\bOmega(\omega)$}
      \label{fig:figB}
    \par\smallskip 
    \vspace{-0.2cm}
\includegraphics[height=3.8cm]{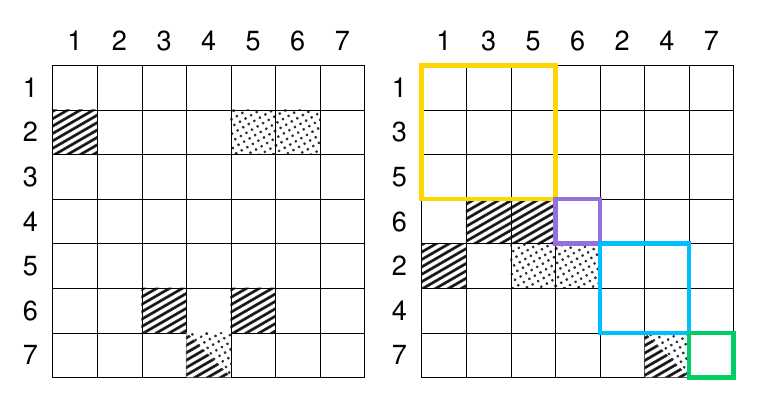}
      \vspace{-0.3cm}
      \caption{$\bA$ (stripe) and $\bB$ (circle)}
      \label{fig:figC}
    \end{subfigure}
  \end{minipage}
  
  \caption{The left panel presents a toy time series chain graph with colors indicating different chain components, and the right panel displays the supports of the original $(\bOmega, \bA, \bB)$ in the first column and the permuted $(\bOmega, \bA, \bB)$ in the second column.}
  \label{fig:three_plots}
\end{figure}

We develop a three-stage learning procedure for recovering the undirected and directed edge sets. 
The first stage optimizes a regularized Whittle likelihood 
with two penalties: a group lasso penalty to enforce group sparsity and a novel tensor-unfolding nuclear norm penalty to capture a common low-rank structure across frequencies. An efficient ADMM algorithm is developed to solve the resulting optimization problem, enabling recovery of the undirected edges.
We then identify the chain components and their causal ordering using a conditional-variance discrepancy measure, and finally estimate the directed edges via multivariate time series regression and thresholding. 
We show theoretically that the proposed procedure consistently recovers the true chain graph structure, and demonstrate its practical effectiveness through extensive simulations and an empirical study of U.S. macroeconomic time series that highlights key features of monetary policy transmission.

Our paper makes useful contributions on multiple fronts. First, we propose a new class of chain graph models for multivariate time series that jointly capture 
contemporaneous and dynamic conditional dependencies within chain components, as well as contemporaneous and dynamic causal relations across chain components. Our formulation yields richer dependence structures and more flexible chain graph modelling compared to chain graph models for independent data \cite[]{zhao2024identifiability} and time-indexed chain graph models for time series \cite[]{dahlhaus2003causality}. Table~\ref{tab:compar} summarizes the comparison.


    \begin{table}[ht]
    \footnotesize	
    \centering
    \caption{\label{tab:compar} The  comparison among three chain graph models.}
    \resizebox{6.5in}{!}
    {\begin{tabular}{lccc}
      \hline
       & \cite{zhao2024identifiability} & \cite{dahlhaus2003causality} & Ours \\
      \hline
      Chain component partitions & Variables & Time indices & Variables \\
      Contemporaneous conditional dependencies & \ding{51} & \ding{51} & \ding{51} \\
      Dynamic conditional dependencies & \ding{55} & \ding{55} & \ding{51} \\
      Contemporaneous causal relations & \ding{51} & \ding{55} & \ding{51} \\
      Dynamic causal relations & \ding{55} & \ding{51} & \ding{51} \\
      \hline
    \end{tabular}}
    \end{table}


On the method side, our proposal involves optimizing a regularized Whittle likelihood with two penalties that simultaneously encourage group sparsity and group low-rank structures shared across frequencies.
The validity of the newly imposed tensor-unfolding nuclear norm penalty for enforcing group low-rank structure is justified through KKT conditions.
To the best of our knowledge, this is the first learning framework that achieves a group sparse plus group low-rank decomposition, which generalizes the well-studied sparse plus low-rank structure \citep{chandrasekaran2011rank} in a groupwise fashion while remaining computationally tractable through an ADMM algorithm. Importantly, the proposed methodology is not restricted to time series chain graph models and can be applied more broadly to settings where a group sparse plus group low-rank decomposition is appropriate, such as 
time series latent variable graphical models \citep{Foti2016} and functional graphical models \citep{Qiao2019} with latent functional variables. 


On the theory side, we are the first to develop an identifiability framework for the group sparse plus group low-rank decomposition by introducing a new transversality condition in the continuous frequency domain. Building on this, we establish a new irrepresentable condition in the same domain, and show that its discretized counterpart, associated with our proposed penalized Whittle likelihood, is satisfied asymptotically, thereby ensuring consistent recovery of both the group sparsity and group low-rank structures. In contrast, \cite{chandrasekaran2011rank} and \cite{zhao2024identifiability} established  theoretical guarantees for the classical sparse plus low-rank decomposition. Our proof involves controlling the discrepancies between continuous and discrete frequencies and employing a novel primal-dual witness technique within the tensor formulation, which provides a suite of technical tools applicable to frequency-domain learning methods for other time series graphical models.


The rest of the paper is organized as follows. 
Section~\ref{sec:model} introduces the time series chain graph formulation and discusses its relationship to relevant work. Section~\ref{sec:method} presents the identifiability in the frequency domain and develops a three-stage learning procedure with an efficient algorithm for recovering the chain graph structure. We establish asymptotic results guaranteeing graph recovery consistency in Section~\ref{sec:theory}. The empirical performance of the proposed method is examined through extensive simulations in Section~\ref{sec:sim} and an application to U.S. macroeconomic data in Section~\ref{sec:real}. 

{\it Notation.} Let $\eZ,\eR^p$ and $\eC^p$ denote the set of integers, the $p$-dimensional real and complex spaces, respectively. For a positive integer $m,$ write $[m]=\{1, \dots, m\}$ and denote by $\bI_m$ the $m \times m$ (complex) identity matrix. 
Let $|c|$ denote the absolute value of a real number $c$, or the modulus of a complex number $c$,  and let $\ti$ denote the imaginary number $\sqrt{-1}.$
For any complex vector $\bz\in\eC^p,\bz^*,\bz^{\H}$ and $\|\bz\|=\sqrt{\bz^{\H}\bz}$ denote its complex conjugate, conjugate transpose and $\ell_2$-norm, respectively.
For a (complex) matrix $\bB=(B_{ij})_{p \times q}$ with singular value decomposition $\sum_{i=1}^{\min(p,q)}\sigma_i\bu_i\bv_i^{\H},$ 
denote its transpose, conjugate transpose, column space, rank and trace (if $\bB$ is a square matrix) by $\bB^{\T},\bB^{\H},\col(\bB),\rank(\bB)$ and $\tr(\bB)$, respectively. 
Denote the operator norm, nuclear norm, Frobenius norm, elementwise $\ell_{\infty}$-norm, and matrix $\ell_1$-norm of $\bB$ by $\|\bB\|=\lambda_{\max}^{1/2}(\bB^{\H}\bB),\|\bB\|_{*}=\sum_{i=1}^{\min(p,q)}\sigma_i,\|\bB\|_{\F}=(\sum_{i,j}|B_{ij}|^2)^{1/2}$, $\|\bB\|_{\max}=\max_{i,j}|B_{ij}|$, and $\|\bB\|_1=\max_{j}\sum_{i}B_{ij}$, respectively, where $\lambda_{\max}(\cdot)$ denotes the largest eigenvalue of a symmetric or Hermitian matrix. 
Additionally, the sub-matrix of $\bB$ corresponding to rows in $S_1$ and columns in $S_2$ is denoted as $\bB_{S_1,S_2}=(B_{ij})_{i\in S_1,j\in S_2},$ and let $\bB_{S_1,S_2}^{-1}$ denote the corresponding sub-matrix of $\bB^{-1}.$
For a vector $\by,$ the sub-vector corresponding to an index subset $S$ is denoted as $\by_S=(\by_i)_{i\in S}.$ 
Let $\cH^p,\cH_+^p$ and $\cH_{++}^p$ denote the set of $p\times p$ Hermitian, Hermitian non-negative definite and Hermitian positive definite complex matrices, respectively. 
We use $\bX\sim N_c(\bmu,\bSigma)$ (or $\bX\sim N_r(\bmu,\bSigma)$) to denote that a complex (or real) random vector $\bX$ follows a complex-valued (or real-valued) multivariate Gaussian distribution. 
For two positive sequences $\{a_n\}$ and $\{b_n\}$, we write $a_n\lesssim b_n$ 
or $b_n\gtrsim a_n$ if there exists a positive constant $c$ such that $a_n/b_n\le c$. 
We write $a_n\asymp b_n$ if and only if $a_n\lesssim b_n$ and $a_n\gtrsim b_n$ hold simultaneously.

\section{Time series Gaussian chain graph model}
\label{sec:model}
\subsection{Model setup}

Suppose that the joint distribution of the strictly and weakly stationary process $\{\bx_t\}_{t\in\eZ}$ 
can be  represented by a time series chain graph $\cG=(\cN,\cE)$, where $\cN=\{1,\dots,p\}$ is  the node set, and the edge set $\cE:=\cE_u\bigcup\cE_d\subset\cN\times\cN$ consists of the undirected and directed edges in $\cE_u$ and $\cE_d$, respectively. 
Let  $(l-k)$ denote an undirected edge between nodes $l$ and $k$, and $(l\to k)$ denote a directed edge from nodes $l$ to $k$.
Assume that at most one edge may exist between any pair of nodes.
For each node $k\in\cN$, define its parent, child, and neighbor sets as $\tpa(k)=\{l\in\cN:(l\to k)\in\cE_d\},\tch(k)=\{l\in\cN:(k\to l)\in\cE_d\}$ and $\tne(k)=\{l\in\cN:(l-k)\in\cE_u\}$,  respectively. 
The node set $\cN$ can then be uniquely partitioned into $G$ disjoint chain components as $\cN=\bigcup_{g=1}^{G}\tau_g$, where each $\tau_g$ forms a 
connected subgraph through undirected edges. 
For a chain component $\tau_g$, we further define its parent set as $\tpa(\tau_g)=\bigcup_{k\in\tau_g}\tpa(k)$. 
We impose two structural assumptions. First,
undirected edges are allowed only within each chain component, whereas directed edges are permitted only between different chain components. 
Second, suppose there exists a permutation $\bpi=(\pi_1,\dots,\pi_G)$ such that, for any $l\in\tau_{\pi_g}$ and $k\in\tau_{\pi_h},$ if $(l\to k)\in\cE_d,$ then $g<h$ \citep{zhao2024identifiability}.
We refer to $\bpi$ as the causal ordering of the chain components. Under this ordering, directed edges point only from higher- to lower-ordered components, thereby ensuring the acyclicity across different chain components.

We consider a $p$-dimensional real-valued time series $\{\bx_t\}$ following
\begin{equation}
    \label{eq:model}
    \bx_t=\bA\bx_t+\bB\bx_{t-1}+\be_t,\quad t\in[T],
\end{equation}
where $\bA=(A_{kl})_{p \times p}$ and $\bB=(B_{kl})_{p \times p}$ are the coefficient matrices capturing, respectively, contemporaneous and dynamic causal relations, and $\{\be_t\}$ is a real-valued, zero-mean stationary Gaussian time series. 
Let $\bSigma_e(h) = \eE(\be_t\be_{t-h}^{\T})$ be the lag-$h$ autocovariance matrix of $\{\be_t\}$ for $h\in\eZ$. Under the condition $\sum_{h\in\eZ}\|\bSigma_e(h)\|<\infty,$ the spectral density matrix of $\{\be_t\}$ at frequency $\omega\in(0,2\pi]$ is $\bbf_e(\omega)=(2\pi)^{-1}\sum_{h\in\eZ}\bSigma_e(h)\exp(-\ti\omega h)$. Let  $\bOmega(\omega)=(\Omega_{kl}(\omega))_{p\times p}:=\bbf_e^{-1}(\omega)$. 
By Proposition 2.2 of \cite{dahlhaus2000graphical}, $\Omega_{kl}(\omega)=0$ for all $\omega\in(0,2\pi]$ if and only if  $\{e_{tk}\}$ and $\{e_{tl}\}$ are conditionally independent given all remaining subprocesses $\{\be_{t,\{k,l\}^c}\}$.  Thus $\{\bOmega(\omega):\omega\in(0,2\pi]\}$
encodes the conditional dependence structure of $\{\be_t\}.$ 
Let $A_{kl}\neq 0$ or $B_{kl}\neq 0$ if and only if $l\in\tpa(k),$ and $\Omega_{kl}(\omega)\neq 0$ for some $\omega\in(0,2\pi]$ if and only if $l\in\tne(k).$ The directed and undirected edges in $\cG$ are then determined by the nonzero entries of $(\bA,\bB)$ and nonzero cross-frequency entries of $\{\bOmega(\omega):\omega\in(0,2\pi]\}$, respectively. Consider the example in Figure~\ref{fig:three_plots}. Within the yellow chain component, $\Omega_{13}(\omega)\neq 0$ and $\Omega_{35}(\omega)\neq 0$ for some $\omega$ correspond to the undirected edges ($1-3$) and $(3-5)$, respectively. Between different chain components, e.g., $A_{36} \neq 0$ represents a contemporaneous directed edge ($3 \rightarrow 6$), $B_{52} \neq 0$ corresponds to a dynamic (lag-1) directed edge ($5 \rightarrow 2$) and $A_{47} \neq 0, B_{47} \neq 0$ indicate a directed edge ($4 \rightarrow 7$) both contemporaneously and dynamically.

Suppose that the joint distribution of $\bx_t$ satisfies the AMP Markov property \citep{andersson2001alternative} with respect to $\cG$.  The density of $\bx_t$ then
admits the factorization
\begin{equation}
    \label{eq:factorization}
    \begin{aligned}
        &\eP(\bx_t)=\prod_{g=1}^{G}\eP(\bx_{t,\tau_g}|\bx_{t,\tpa(\tau_g)},\bx_{t-1,\tpa(\tau_g)}),\\
        &\bx_{t,\tau_g}|\bx_{t,\tpa(\tau_g)},\bx_{t-1,\tpa(\tau_g)}\sim N_r\big(\bA_{\tau_g,\tpa(\tau_g)}\bx_{t,\tpa(\tau_g)}+\bB_{\tau_g,\tpa(\tau_g)}\bx_{t-1,\tpa(\tau_g)},\bSigma_{e,\tau_g,\tau_g}(0)\big).
    \end{aligned}
\end{equation}
Furthermore, conditional on $\{\bx_{t,\tpa(\tau_g)}\}_{t\in\eZ}$, the lag-$h$ autocovariance of $\{\bx_{t,\tau_g}\}_{t\in\eZ}$ equals $\bSigma_{e,\tau_g,\tau_g}(h)$ for $h \in \mathbb{Z}$, and the conditional dependence structure of $\{\bx_{t,\tau_g}\}_{t\in\eZ}$ coincides with that of $\{\be_{t,\tau_g}\}_{t\in\eZ}$.
Hence, for each chain component $\tau_g$,  given its parent set $\tpa(\tau_g)$, both the distribution and the conditional dependence structure of $\{\bx_{t,\tau_g}\}$ are fully characterized by those of $\{\be_{t,\tau_g}\}$. This, in turn, justifies the construction of $\cG$. To further ensure the acyclicity across chain components in $\cG$,  we define $(\bOmega,\bA,\bB)$ to be time series chain graph (TSCG)-feasible. 
\begin{definition}
    \label{def:CG}
    A triplet $(\bOmega,\bA,\bB)$ is TSCG-feasible if there exists a permutation matrix $\bP\in\eR^{p\times p}$ such that $\{\bP\bOmega(\omega)\bP^{\T}:\omega\in(0,2\pi]\},$ $\bP\bA\bP^{\T}$ and $\bP\bB\bP^{\T}$ share the same block structure, where $\bP\bOmega(\omega)\bP^{\T}$ is a block diagonal matrix for each $\omega\in(0,2\pi]$, and $\bP\bA\bP^{\T}$ and $\bP\bB\bP^{\T}$ are block lower triangular matrices with zero diagonal blocks. 
\end{definition}
Figure~\ref{fig:three_plots} provides an illustrative example of the common block structure described in Definition~\ref{def:CG}, obtained via appropriate row and column permutations of $(\bOmega,\bA,\bB).$

\begin{remark}
Model~\eqref{eq:model} admits a natural extension to accommodate $d$ lags of $\bx_t$:
\begin{equation}
\label{TSCG.extend}
\bx_t=\bA\bx_t+\sum_{h=1}^{d}\bB_h\bx_{t-h}+\be_t,
\end{equation}
where $\bA$ represents the contemporaneous causal relations,  and $\bB_h$ encodes the dynamic causal relations at lag $h\in[d]$.  Both the identifiability results in Section~\ref{subsec:iden} and the graph learning algorithm in Section~\ref{subsec:alg} can be extended accordingly to this general setting.
For ease of exposition, however, we focus on the case $d = 1.$ See Remark~\ref{rmk:lagd} for further discussion.
\end{remark}

\begin{remark}
    \label{rmk:e_var}

    To gain further insight into the dependence structure of the CIG within each chain component $\tau_g$ for $g \in [G]$, one may rely on the Granger causality graph for identifying the contemporaneous conditional dependencies and dynamic Granger-causal relations within each $\tau_g$. Following \citet{Eichler2007}, we associate the Granger causality graph of $\{\be_t\}$ with a VAR representation. For simplicity, suppose that $\{\be_t\}$ follows a VAR(1) model as
    $\be_t=\bC\be_{t-1}+\bvarepsilon_t,$ where $\{\bvarepsilon_t\}$ is a white noise sequence with covariance matrix $\bSigma_{\varepsilon}$. Under the TSCG-feasibility in Definition~\ref{def:CG}, $\{\be_{t,\tau_g}\}$ and $\{\be_{t,\tau_{g'}}\}$ are independent for any $g\neq g',$ and  up to a permutation, the matrices $\bC$ and $\bSigma_{\varepsilon}^{-1}$ share the same block-diagonal structure as $\bOmega(\omega)$ for $\omega\in(0,2\pi]$.  Then, each $\tau_g$ admits a separate VAR representation:
    \begin{equation}
        \label{eq:var_cau}
        \be_{t,\tau_g}=\bC_{\tau_g,\tau_g}\be_{t-1,\tau_g}+\bvarepsilon_{t,\tau_g},\quad\bvarepsilon_{t,\tau_g}\sim N_r(\bzero,\bSigma_{\varepsilon,\tau_g,\tau_g}),\quad g\in[G].
    \end{equation}
The directed and undirected edges in the Granger causality graph thus correspond to the nonzero entries of the coefficient matrix $\bC_{\tau_g,\tau_g}$ and the precision matrix $\bSigma_{\varepsilon,\tau_g,\tau_g}^{-1}$, respectively.


To illustrate, we consider the yellow chain component $\tau_1=\{1,3,5\}$ in Figure~\ref{fig:three_plots}. Suppose that $\be_{t,\tau_1}$ follows a VAR(1) model
\begin{equation} \label{model:var}
    \be_{t,\tau_1} = \bC_{\tau_1,\tau_1} \be_{t-1,\tau_1} + \bvarepsilon_{t,\tau_1}, \quad \bvarepsilon_{t,\tau_1} \sim \mathcal{N}_r(\bzero, \bSigma_{\varepsilon,\tau_1,\tau_1}),
\end{equation}
where 
\vspace{-1.0em}
{\small
$$
\begingroup
\renewcommand{\arraystretch}{0.7}
\setlength{\arraycolsep}{3.5pt}
\bC_{\tau_1,\tau_1} = \begin{pmatrix}
0.6 & 0.2 & 0\\
0 & 0.6 & 0\\
0 & 0 & 0.6
\end{pmatrix}, \quad
\bSigma_{\varepsilon,\tau_1,\tau_1} = \begin{pmatrix}
1.0 & 0 & 0\\
0 & 1.0 & 0.5\\
0 & 0.5 & 1.0
\end{pmatrix}.
\endgroup
$$
}Figure~\ref{fig:causality} presents the corresponding conditional independence and Granger causality graphs, where the undirected edges in Figure~\ref{fig:2.1} are determined by nonzero cross-frequency entries of $\{\bOmega_{\tau_1,\tau_1}(\omega):\omega\in(0,2\pi]\}$, and
the directed and undirected edges in Figure~\ref{fig:2.2} are determined by the nonzero entries of $\bC_{\tau_1,\tau_1}$ and $\bSigma_{\varepsilon,\tau_1,\tau_1}^{-1}$, respectively. 
It is noteworthy that we implicitly assume that each component series depends on its own past \citep{Eichler2007}, which could be represented by directed self-loops. Since the insertion of these loops does not affect the separation properties for the Granger causality graphs, we omit them for simplicity.
See Remark~\ref{rm.CIG} and the real-data application in Section~\ref{sec:real} for further details on the estimation of the Granger causality graph subject to the CIG and its implementation.

    \begin{figure}[ht]
  \centering
  
   \begin{minipage}{0.45\linewidth}
   \hspace{3em}
       \begin{subfigure}{\linewidth}
      \centering
\includegraphics[height=2.3cm]{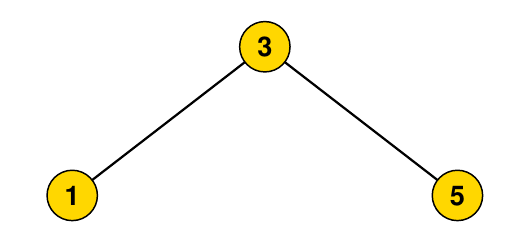}
      \vspace{-0.1cm}
      \caption{Conditional independence graph} \label{fig:2.1}
    \end{subfigure}
    \end{minipage}
  \hfill
  \begin{minipage}{0.45\linewidth}
    \hspace{-3em}
    \begin{subfigure}{\linewidth}
      \centering
\includegraphics[height=2.3cm]{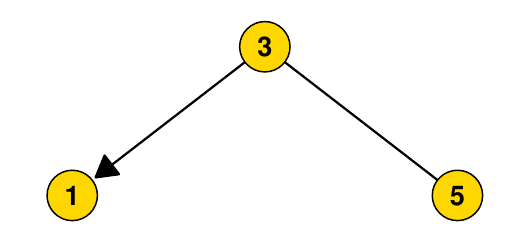}
      \vspace{-0.1cm}
      \caption{Granger causality graph} \label{fig:2.2}
    \end{subfigure}
  \end{minipage} 
  \caption{Conditional independence and Granger causality graphs for the yellow chain component in Figure~\ref{fig:three_plots} under model~\eqref{model:var}.}
  \label{fig:causality}
\end{figure}
\end{remark}


\subsection{Relationship to relevant work}

Model~\eqref{eq:model} jointly captures  temporal and cross-sectional dependence structures  and is related to several multivariate time series models in the literature. 
First, consider the case $\bA=\bzero$, $\bB\neq\bzero.$ If $\{\be_t\}$ is white noise, then model~\eqref{eq:model} reduces to the standard VAR model, 
which was also used by \citet{dahlhaus2003causality} for the time-indexed chain graph model and by \citet{Eichler2007} for the Granger causality graph.
If $\be_t=\bH_t^{1/2}\bfeta_t$, where $\bH_t$ denotes the conditional covariance matrix of $\bx_t$ given past information, and $\{\bfeta_t\}$ is i.i.d. innovations with zero mean and identity covariance matrix, then model~\eqref{eq:model} can be written as $\bx_t=\bB\bx_{t-1}+\bH_t^{1/2}\bfeta_t,$ which corresponds to a vector AR-GARCH model \citep[e.g.,][]{ling2003asymptotic}. 
Under suitable regularity conditions, the process $\be_t=\bH_t^{1/2}\bfeta_t$ is unconditionally stationary (both strictly and weakly), as assumed in our framework.

Second, consider the case $\bA\neq\bzero$ and $\bB\neq\bzero.$ If $\{\be_t\}$ is white noise with a diagonal covariance matrix, model~\eqref{eq:model} can be viewed as a structural VAR (SVAR) model \citep{sims1980macroeconomics}, i.e.,  $\bA_*\bx_t=\bB\bx_{t-1}+\be_t,$ or equivalently, $\bx_t=\bA_*^{-1}\bB\bx_{t-1}+\bu_t,$ where $\bA_*=\bI_p-\bA$ is an invertible structural coefficient matrix that captures contemporaneous relations, and $\bu_t=\bA_*^{-1}\be_t$ is the reduced-form residual. 
Identification of 
SVAR typically requires restrictions on $\bA_*$ (or its inverse), and a common choice is the short-run restriction \citep[see Section 9.1.1 of][]{lutkepohl2005new},
 which specifies $\bA_*$ to be lower-triangular. 
 Recall that $(\bOmega,\bA,\bB)$ is TSCG-feasible in Definition~\ref{def:CG}. Without loss of generality, we can assume $\bA$ is block lower-triangular, which implies that $\bA_*=\bI_p-\bA$ is lower-triangular and is consistent with the short-run identification restriction.
Model~\eqref{eq:model} is also related to spatio-temporal autoregressive (STAR) models \citep[e.g.,][]{gao2019banded,ma2023sparse}, where $\bA$ and $\bB$ capture the spatial and temporal dependencies, respectively. A key difference is that classical STAR models typically assume i.i.d. innovations $\{\be_t\}$, whereas our framework allows for both temporal and cross-sectional dependencies in $\{\be_t\}$, yielding a richer dependence structure.

Beyond the aforementioned multivariate time series models, our formulation coincides with the linear structural equation model \citep{peters2014identifiability,park2020identifiability} when $\bA\neq\bzero,\bB=\bzero$ and $\{\be_t\}$ is an i.i.d. sequence, and further relates to
the chain graph model of \citet{zhao2024identifiability} for independent data, where 
the causal relations and conditional dependencies among nodes are respectively represented by the nonzero entries of $\bA$ and $\bSigma_e^{-1}(0)$. Compared with our proposed time series chain graph model, a direct application of 
the formulation in \citet{zhao2024identifiability} to the time series setting fails to capture the temporal dependence structure encoded in the lagged coefficient matrix $\bB$ and the inverse spectral density matrix $\bOmega(\omega)$, and may therefore lead to spurious detected edges. To illustrate, consider  an example of $3$-dimensional time series $\{\bx_t\}$ satisfying
$\bx_t=\bA\bx_t+\be_t$ for $t\in[T],$
where the error process follows $\be_t=\alpha \bP \bw_{t-1} + \bw_t$ with $|\alpha| < 1$,  $\bP$ is a $3 \times 3$ cyclic permutation matrix and $\{\bw_t\}$ is a white noise sequence. In this example, although $\bB = 0$ and both models capture the same set of directed edges implied by $\bA$,  $\{\be_t\}$ still exhibits temporal dependence that cannot be characterized by the covariance-based formulation of \citet{zhao2024identifiability}.
Specifically, since $\bSigma_e^{-1}(0) = (1+\alpha^2)\bI_3$ is diagonal, their model incorrectly identifies no undirected edges. In contrast, our model can correctly capture the conditional dependencies between all three pairs, i.e., $(1-2),(2-3)$ and $(1-3)$, through $\{\bOmega(\omega):\omega\in(0,2\pi]\}$.

\section{Methodology}
\label{sec:method}

\subsection{Identifiability}
\label{subsec:iden}
Let $\bbf_x(\omega)$ denote the spectral density matrix of $\bx_t$ at frequency $\omega\in (0,2\pi]$. 
Under model \eqref{eq:model} with $\be_t=(\bI_p-\bA)\bx_t-\bB\bx_{t-1}$,  we can write
$\bbf_e(\omega)=(\bI_p-\bA-\bB\exp\big(-\ti\omega)\big)\bbf_x(\omega)\big(\bI_p-\bA-\bB\exp(-\ti\omega)\big)^{\H}.$ 
We define the inverse spectral density matrix of $\bx_t$ by $\bTheta(\omega)=\bbf_x^{-1}(\omega)\in\cH_{++}^p$. 
It then follows from the TSCG-feasibility of $(\bOmega,\bA,\bB)$ in Definition~\ref{def:CG} that
\begin{equation}
    \label{eq:Omega}
    \bTheta(\omega)=\big(\bI_p-\bA-\bB\exp(-\ti\omega)\big)^{\H}\bOmega(\omega)\big(\bI_{p} - \bA - \bB\exp(-\ti\omega)\big) =: \bOmega(\omega) + \bL(\omega),
\end{equation}
where $\bL(\omega):=\big(\bA+\bB\exp(-\ti\omega)\big)^{\H}\bOmega(\omega)\big(\bA+\bB\exp(-\ti\omega)\big)-\big(\bA+\bB\exp(-\ti\omega)\big)^{\H}\bOmega(\omega)-\bOmega(\omega)\big(\bA+\bB\exp(-\ti\omega)\big)$ with $\bL(\omega)\in\cH_+^p$. 

 
 Motivated further by Definition~\ref{def:CG},
 we assume that $\bOmega(\omega)$ is group sparse across $\omega\in(0,2\pi]$, which implies the sparseness of undirected edges in $\cG$ and has been well adopted in the literature of time series Gaussian graphical models \cite[]{jung2015graphical,tugnait2022sparse}.
 We also assume that $\bL(\omega)$ is group low-rank across $\omega\in(0,2\pi]$, arising naturally from the low-rank structures of the coefficient matrices $\bA$ and $\bB$ that essentially capture the presence of hub nodes in $\cG$ \citep{fang2023low}. 
 Specifically, the group support of $\bOmega(\cdot)$ and the group rank of $\bL(\cdot)$ are respectively defined as:
$$
\gsupp(\bOmega):=\{(k,l):\Omega_{kl}(\omega)\neq0,\exists\,\omega\in(0,2\pi]\},\quad\grank(\bL):=\sup_{\omega\in(0,2\pi]}\rank\big(\bL(\omega)\big),
$$ 
with $|\gsupp(\bOmega)|=S<p^2$ and $\grank(\bL)=R<p$, where $|\cdot|$ denotes the cardinality of a set. 


For each frequency $\omega\in(0,2\pi]$, consider the eigen-decomposition $\bL(\omega)=\bU\bD(\omega)\bU^{\H}$, where $\bU^{\H}\bU=\bI_R$ and $\bD(\omega)$ is a $R\times R$ real-valued diagonal matrix. Let $\cC(\cX;\cY)$ denote the set of continuous functions mapping the domain $\cX$ to the codomain $\cY.$ We then introduce two linear subspaces in $\cC((0,2\pi];\cH^p)$: 
$$
\begin{aligned}
    \cS(\bOmega)=&\big\{\bOmega'\in\cC((0,2\pi];\cH^p):\gsupp(\bOmega')\subset\gsupp(\bOmega)\big\},\\
    \cT(\bL)=&\big\{\bL'\in\cC((0,2\pi];\cH^p):\bL'(\omega)=\bU\bV(\omega)+\bV(\omega)^{\H}\bU^{\H}\ {\rm for\ some\ }\bV(\omega)\in\eC^{R\times p}\big\},
\end{aligned}
$$
where $\cS(\bOmega)$ is the tangent space at point $\bOmega$ with respect to the algebraic variety defined as $\{\bOmega'\in\cC((0,2\pi];\cH^p):|\gsupp(\bOmega')|\le S\}$, and $\cT(\bL)$ is the tangent space at point $\bL$ with respect to the algebraic variety defined as $\{\bL'\in\cC((0,2\pi];\cH^p):\grank(\bL')\le R\}$. 
Let $(\bOmega_0,\bA_0,\bB_0)$ denote the true parameters of model~\eqref{eq:model}, and $\bL_0(\cdot)$ be the true value of $\bL(\cdot)$ with $R_0:=\grank(\bL_0)$. 
To ensure the identifiability of the time series chain graph $\cG$, we impose the following assumptions. 

\begin{assumption}
    \label{ass:origin}
    $\cS(\bOmega_0)\bigcap\cT(\bL_0)=\{\bzero(\cdot)\}$, where $\bzero(\cdot)$ is the origin of the space $\cC((0,2\pi];\cH^p)$ such that $\bzero(\omega)=\bzero_{p\times p}$ for all $\omega\in(0,2\pi]$.
\end{assumption}

\begin{assumption}
    \label{ass:dis_eigen}
    The $R_0$ eigenvalues of $\bL_0(\omega)$ are distinct for $\omega\in(0,2\pi].$
\end{assumption}

The transversality condition in Assumption~\ref{ass:origin} ensures that the tangent spaces $\cS(\bOmega_0)$ and $\cT(\bL_0)$ intersect only at the origin \citep{chandrasekaran2011rank,chandrasekaran2012latent}. This  guarantees the unique decomposition of  $\bTheta_0\in\cC((0,2\pi];\cH^p)$ into the sum of one function in $\cS(\bOmega_0)$ and another in $\cT(\bL_0)$, where $\bTheta_0(\cdot)$ denotes the true value of $\bTheta(\cdot).$ It essentially requires that the group sparse  $\bOmega_0(\cdot)$ is not group low-rank, and the group low-rank  $\bL_0(\cdot)$ is not group sparse.
In our framework,  the inverse spectral density matrix $\bOmega_0(\omega)$ is  full-rank for $\omega \in (0,2\pi]$ and
$\bL_0(\omega)$ is generally non-sparse as each of its entries relates to interactions among multiple nodes through $(\bOmega_0,\bA_0,\bB_0)$. Specifically,
the $(k,l)$-th entry of $\bL_0(\omega)$ is given by $L_{0,kl}(\omega)=\sum_{i=1}^{p}\sum_{j=1}^{p}\big(A_{0,ik}+B_{0,ik}\exp(\ti\omega)\big)\Omega_{0,ij}(\omega)\big(A_{0,jl}+B_{0,jl}\exp(-\ti\omega)\big)-\sum_{i=1}^{p}\big(A_{0,ik}+B_{0,ik}\exp(\ti\omega)\big)\Omega_{0,il}(\omega)-\sum_{j=1}^{p}\Omega_{0,kj}(\omega)\big(A_{0,jl}+B_{0,jl}\exp(-\ti\omega)\big)$, where the three terms correspond to the path $k\to i-j\gets l$ if $i\neq j$ or $k\to i\gets l$ if $i=j$, the path $k\to i-l$, and the path $k-j\gets l$, respectively. Hence, $L_{0,kl}(\omega)\neq0$ if any such path exists between nodes $k$ and $l$.
Assumption~\ref{ass:dis_eigen} is standard in the matrix perturbation literature \citep{yu2015useful} to ensure the identifiability of the eigenspace of the low-rank matrix $\bL_0(\omega)$ for $\omega\in(0,2\pi]$.

\begin{assumption}
    \label{ass:stationary}
    $\{\be_t\}_{t\in\eZ}$ is strictly stationary and ergodic, and $\rho\{(\bI_p-\bA)^{-1}\bB\}<1$, where $\rho(\cdot)$ denotes the spectral radius of a matrix.
\end{assumption}

\begin{assumption}
    \label{ass:cond_mean}
    $\eE\big(\be_{t,\tau_g}|\cF_{t,\tpa(\tau_g)}\big)=\bzero$ for $g\in[G],$ where $\cF_{t,\tpa(\tau_g)}=\sigma(\bx_{t,\tpa(\tau_g)},\bx_{t-1,\tpa(\tau_g)},\dots).$
\end{assumption}

Assumption~\ref{ass:stationary} ensures that $\bx_t$ is both strictly and weakly stationary under model~\eqref{eq:model}. To see this, note that model~\eqref{eq:model} can be rewritten as $\bx_t=(\bI_p-\bA)^{-1}\bB\bx_{t-1}+(\bI_p-\bA)^{-1}\be_t$, where $\bI_p -\bA$ is always invertible due to the acyclicity across chain components in $\cG$. The stationarity then follows directly from Theorem A.1 of \citet{francq2019garch}.
Assumption~\ref{ass:cond_mean} is known as the weak exogeneity condition in the literature \citep[e.g.,][]{mikusheva2025linear}.
It implies that $\eE(\bx_{t,\tau_g}|\bx_{t,\tpa(\tau_g)},\bx_{t-1,\tpa(\tau_g)})=\bA_{\tau_g,\tpa(\tau_g)}\bx_{t,\tpa(\tau_g)}+\bB_{\tau_g,\tpa(\tau_g)}\bx_{t-1,\tpa(\tau_g)}$ for $g\in[G],$ which is necessary for the identifiability of $\bA$ and $\bB.$ 

Let $\cQ$ denote the parameter space of TSCG-feasible triplets $(\bOmega,\bA,\bB)$, where $\bOmega(\cdot)\in\cC((0,2\pi];\cH_{++}^p)$, $\bL(\cdot)\in\cC((0,2\pi];\cH_+^p),|\gsupp(\bOmega)|\le |\gsupp(\bOmega_0)|$, and $\grank(\bL)\le R_0$. Write $\vvvert\bOmega\vvvert_{\max}=\sup_{\omega\in(0,2\pi]}\|\bOmega(\omega)\|_{\max}$.


\begin{theorem}
    \label{thm:iden}
    Suppose that Assumptions~\ref{ass:origin}--\ref{ass:cond_mean} hold. Then, there exists a small $\epsilon>0$ such that for any $(\bOmega,\bA,\bB)\in\cQ$ satisfying $\vvvert\bOmega-\bOmega_0\vvvert_{\max}<\epsilon,\|\bA-\bA_0\|_{\max}<\epsilon$ and $\|\bB-\bB_0\|_{\max}<\epsilon,$ if $\big(\bI_p-\bA-\bB\exp(-\ti\omega)\big)^{\H} \bOmega(\omega)\big(\bI_p-\bA-\bB\exp(-\ti\omega)\big)=\big(\bI_p-\bA_0-\bB_0\exp(-\ti\omega)\big)^{\H}\bOmega_0(\omega)\big(\bI_p-\bA_0-\bB_0\exp(-\ti\omega)\big)$ for $\omega\in(0,2\pi]$, then $(\bOmega,\bA,\bB)=(\bOmega_0,\bA_0,\bB_0).$
\end{theorem}

Theorem~\ref{thm:iden} shows that the time series chain graph $\cG$ is locally identifiable. Specifically,  the true cross-frequency inverse spectral density matrix $\{\bTheta_0(\omega):\omega\in(0,2\pi]\}$ uniquely determines the true parameter triplet $(\bOmega_0,\bA_0,\bB_0)$ within its neighborhood in $\cQ$.


\subsection{Estimation procedure}
\label{subsec:alg}
 The discrete Fourier transform (DFT) 
 serves as a key tool in our methodological development, as it transforms the temporally dependent sequence $\{\bx_t\}$ into an approximately independent Gaussian sequence.  
Define the (normalized) DFT of $\{\bx_t\}$ as
\begin{equation}
    \label{eq:dft}
    \bd_x(\omega_j) = \frac{1}{\sqrt{T}}\sum_{t=1}^{T}\bx_t \exp(-\ti\omega_jt),
\end{equation}
where $\omega_j=2\pi j/T$ for $j\in[T]$ is the Fourier frequency. Without loss of generality, we assume that $T$ is even. 
By Theorem 4.4.1 of \cite{brillinger2001time}, as $T\to\infty$, $\bd_x(\omega_j)$ for $j\in [T/2-1]$ are independent circularly symmetric complex-valued Gaussian with $N_c\big(\bzero, 2\pi\bbf_x(\omega_j)\big),$ and $\bd_x(\omega_j)$ for $j\in\{T/2,T\}$ are independent real-valued Gaussian with $N_r\big(\bzero,2\pi\bbf_x(\omega_j)\big).$ 
Since $T^{-1/2}\sum_{t=0}^{T-1}\bx_t\exp(-\ti\omega_jt)-\exp(-\ti\omega_j)\bd_x(\omega_j)=T^{-1/2}(\bx_0-\bx_T)=o_p(1),$ we can rewrite model~\eqref{eq:model} in the frequency domain as
\begin{equation}
    \label{eq:model_dx}
    \bd_x(\omega_j)=\big\{\bA+\bB\exp(-\ti\omega_j)\big\}\bd_x(\omega_j)+\bd_e(\omega_j)+o_p(1),
\end{equation}
which further implies that, for $g\in[G]$ and $j\in[T]$, as $T\to\infty$,
{\small\begin{equation}
    \label{eq:cond_var_d}
    \bd_{x,\tau_g}(\omega_j) | \bd_{x,\tpa(\tau_g)}(\omega_j)\to_d N_c\Big(\{\bA_{\tau_g, \tpa(\tau_g)}+\bB_{\tau_g,\tpa(\tau_g)}\exp(-\ti\omega_j)\}\bd_{x,\tpa(\tau_g)}(\omega_j),2\pi\bOmega_{\tau_g,\tau_g}^{-1}(\omega_j)\Big).
\end{equation}}

\begin{remark} 
\label{rmk:lagd}
For the generalization of model~\eqref{eq:model} with $d$ lags in \eqref{TSCG.extend}, the corresponding frequency-domain representation is $\bd_x(\omega_j)=\big\{\bA+\sum_{h=1}^{d}\bB_h\exp(-\ti h\omega_j)\big\}\bd_x(\omega_j)+\bd_e(\omega_j)+o_p(1)=:\widetilde{\bA}(\omega_j)\bd_x(\omega_j)+\bd_e(\omega_j)+o_p(1)$ for $j\in[T]$, where 
 $\widetilde{\bA}(\omega_j)$ represents the coefficient matrix in the frequency domain. The identifiability results  in Section~\ref{subsec:iden} and the estimation procedure below can then be naturally extended to accommodate this lag-$d$ formulation.
\end{remark}

We next develop a three-stage procedure for recovering the time series chain graph $\cG.$ 

The first stage estimates the undirected edge set $\cE_u$ through the group sparsity structure of $\bOmega(\cdot)$. Since 
$\bd_x(\omega_j)^{*}=\bd_x(-\omega_j)=\bd_x(2\pi-\omega_j)$, we focus on non-negative Fourier frequencies. Recall that $\bd_x(\omega_j)$ are asymptotically independent 
$N_c\big(\bzero,2\pi\bbf_x(\omega_j)\big)$ for $j\in [T/2-1]$. Then, the log-likelihood function (ignoring the constants) can be approximated as:
\begin{equation}
    \label{eq:likelihood}
    \begin{aligned}
        l(\bTheta)\approx&\sum_{j=1}^{T/2-1} \big[\log \det\{\bTheta(\omega_j)\} -(2\pi)^{-1}\tr\{\bTheta(\omega_j)\bd_x(\omega_j)\bd_x(\omega_j)^{\H}\}\big]\\
        =&\sum_{j=1}^{M}\sum_{n=-m}^{m}\big[\log \det\{\bTheta(\widetilde{\omega}_{j,n})\} -(2\pi)^{-1}\tr\{\bTheta(\widetilde{\omega}_{j,n})\bd_x(\widetilde{\omega}_{j,n})\bd_x(\widetilde{\omega}_{j,n})^{\H}\}\big]\\
        \approx&(2m+1)\sum_{j=1}^{M}\big[\log \det\{\bTheta(\widetilde{\omega}_j)\} -\tr\{\bTheta(\widetilde{\omega}_j)\widehat{\bbf}_x(\widetilde{\omega}_j)\}\big],
    \end{aligned}
\end{equation}
where $m$ denotes the pre-specified half-block size, 
$M=\lfloor (T/2-1)/(2m+1)\rfloor$ is the number of equally spaced frequency blocks, $\widetilde{\omega}_{j,n}=\omega_{j(2m+1)-m+n}$ for $j\in[M],n\in\{-m,-(m-1),\dots,m\}$ are Fourier frequencies, $\widetilde{\omega}_j=\widetilde{\omega}_{j,0}$ is the central frequency in the $j$-th block to evaluate the log-likelihood function, and the estimated spectral density matrix of $\bx_t$ at frequency $\widetilde{\omega}_j$ is
\begin{equation}
    \label{eq:spec_est}
    \widehat{\bbf}_x(\widetilde{\omega}_j)=\frac{1}{2\pi(2m+1)}\sum_{n=-m}^{m}\bd_x(\widetilde{\omega}_{j,n})\bd_x(\widetilde{\omega}_{j,n})^{\H}=\frac{1}{2m+1}\sum_{n=-m}^{m}\bI_x(\widetilde{\omega}_{j,n}).
\end{equation}
Here $\bI_x(\widetilde{\omega}_{j,n})=(2\pi)^{-1}\bd_x(\widetilde{\omega}_{j,n})\bd_x(\widetilde{\omega}_{j,n})^{\H}$ denotes the periodogram.  While a single periodogram $\bI_x(\widetilde{\omega}_j)$ is an   inconsistent estimator of $\bbf_x(\widetilde{\omega}_j),$
we average $2m+1$ periodograms over consecutive frequencies to obtain a consistent estimator as $m \rightarrow \infty$ with $T \rightarrow \infty.$ 
In \eqref{eq:likelihood}, the first ``$\approx$'' follows from the Whittle likelihood with theoretical guarantees provided by Theorem 10.3.2 of \cite{brockwell1991time},
and the second ``$\approx$'' follows from the local smoothness of the spectral density matrix \citep[see Theorem 10.4.1 of][]{brockwell1991time}, which implies that $\bbf_x(\widetilde{\omega}_{j,0})\approx\bbf_x(\widetilde{\omega}_{j,n})$ for $n=-m,-m+1,\dots,m$.

For simplicity, denote by $\widetilde{\bTheta},\widetilde{\bOmega},\widetilde{\bL}\in\eC^{p\times p\times M}$ the order-3 complex-valued tensors with slices $\widetilde{\bTheta}_{::j}=\bTheta(\widetilde{\omega}_j),\widetilde{\bOmega}_{::j}=\bOmega(\widetilde{\omega}_j),\widetilde{\bL}_{::j}=\bL(\widetilde{\omega}_j)$ for $j\in[M],$ where $\widetilde{\bTheta}_{::j}$ denotes the mode-3 slice of $\widetilde{\bTheta}$ indexed by $(:,:,j)$, and similarly for $\widetilde{\bOmega}_{::j}$ and $\widetilde{\bL}_{::j}$. Define the Whittle log-likelihood approximation as $\ell_M(\widetilde{\bTheta})=\sum_{j=1}^{M}[\log \det\{\bTheta(\widetilde{\omega}_j)\} -\tr\{\bTheta(\widetilde{\omega}_j)\widehat{\bbf}_x(\widetilde{\omega}_j)\}].$ We estimate  $\widetilde{\bOmega}$ by minimizing the following regularized Whittle likelihood:
\begin{equation}
    \label{eq:min}
    \begin{aligned}
        (\widehat{\bOmega},\widehat{\bL})=& \argmin_{\widetilde{\bOmega},\widetilde{\bL}} -\ell_M(\widetilde{\bOmega}+\widetilde{\bL}) + P_1(\widetilde{\bOmega},\lambda_{1T})+P_2(\widetilde{\bL},\lambda_{2T}),\\
        &\text{s.t.}\quad\bOmega(\widetilde{\omega}_j)\succ\bzero,\quad\bL(\widetilde{\omega}_j)\succcurlyeq\bzero,\quad j\in[M],
    \end{aligned}
\end{equation}
where $P_1(\widetilde{\bOmega},\lambda_{1T})=\lambda_{1T}\sqrt{M}\sum_{k\neq l}\sqrt{\sum_{j=1}^{M}|\Omega_{kl}(\widetilde{\omega}_j)|^2}$ is the group lasso penalty \citep{yuan2006model} to enforce group sparsity in $\widetilde{\bOmega}$ across $M$ frequencies with tuning parameter $\lambda_{1T}>0$. 
\begin{equation}
    \label{eq:penalty}
P_2(\widetilde{\bL},\lambda_{2T})=\lambda_{2T}\sqrt{M}\big(\|\widetilde{\bL}_{(1)}\|_{*}+\|\widetilde{\bL}_{(2)}\|_{*}\big)/2
\end{equation}
is a new tensor-unfolding nuclear norm penalty to induce group low-rank of $\widetilde{\bL}$ across $M$ frequencies with tuning parameter $\lambda_{2T}>0,$ where $\widetilde{\bL}_{(q)}$ is the mode-$q$ unfolding of $\widetilde{\bL}$ for $q\in [2].$ The constraints are due to the positive-definiteness of  $\bOmega(\widetilde{\omega}_j)$ and $\bTheta(\widetilde{\omega}_j)$ for $j \in [M].$ Such an optimization problem can be efficiently solved via the ADMM algorithm \cite[]{boyd2011distributed}. See details in Section~\ref{subsec:admm}. 
We then obtain the estimated undirected edge set as  $\widehat{\cE}_u=\big\{(l-k)\in\cN\times\cN:\exists j\in[M],\widehat{\Omega}_{kl}(\widetilde{\omega}_j)\neq0\big\},$ where $\widehat{\bOmega}(\widetilde{\omega}_j)=\widehat{\bOmega}_{::j}$ for $j\in[M].$

\begin{remark}
    \label{rmk:group_rank}
    (i) By the definition of $\bL(\omega)$ and $\bOmega(\omega)\in\cH_{++}^p$, 
    it follows that $\col\big(\bL(\omega)\big)=\col(\bA)\bigcup\col(\bB)\bigcup\col(\bA^{\T})\bigcup\col(\bB^{\T})$ for $\omega\in(0,2\pi],$ and thus 
    $\widetilde{\bL}_{::j}$ shares the same column space for $j\in[M]$, which coincides with that of $\widetilde{\bL}_{(1)}$ and $\widetilde{\bL}_{(2)}$.
    This implies that $\widetilde{\bL}$ exhibits not only the group low-rank structure as defined in Section~\ref{subsec:iden}, but also a stronger common column space structure across its mode-3 slices associated with $M$ frequencies.\\
    (ii) Compared with imposing separate nuclear norm penalties on individual mode-3 slices of $\widetilde{\bL}$ through $\sum_{j=1}^{M}\|\widetilde{\bL}_{::j}\|_{*}$ as in \cite{Foti2016}, our group penalty in \eqref{eq:penalty} better aggregates the column space information across $M$ slices (i.e., $M$ frequencies). 
    Specifically, as implied by the KKT conditions of our optimization problem (see ({\color{blue}S.9}) and ({\color{blue}S.10}) of the supplementary material),
    our tensor-unfolding nuclear norm penalty pools the eigenvalues associated with the common eigenvectors across slices, thus capturing the common column space structure more effectively.
In finite samples, separate penalties may fail to yield the common column space across slices, 
whereas the proposed group penalty provides this guarantee.
\end{remark}

\begin{remark}
\label{rmk.5}
The proposed group sparse plus group low-rank learning framework is general and can be applied broadly to other settings. 
For instance, compared to the method of \citet{Foti2016} for estimating latent variable graphical models of multivariate time series, it provides a more efficient approach, as discussed in Remark~\ref{rmk:group_rank}(ii).
Additionally, our method can be used to estimate functional graphical models \cite[]{Qiao2019} with latent functional variables for multivariate functional data. In a similar spirit to \cite{chandrasekaran2012latent}, this task is equivalent to recovering the group sparse plus group low-rank structure in the precision matrix of the truncated functional principal component (FPC) scores after performing FPC analysis on each functional variable. Hence, our proposed learning framework becomes applicable.
\end{remark}

In the second stage, based on $\widehat{\cE}_u$, we group $p$ nodes into $\widehat G$ estimated chain components $\{\widehat{\tau}_1,\dots,\widehat{\tau}_{\widehat{G}}\}$, and recover the causal ordering of the chain components via an iterative top-down search procedure.
The idea follows from the AMP interpretation of model \eqref{eq:model_dx}, which suggests that the asymptotic conditional variance of node $k$ given its all parent nodes at frequency $\widetilde{\omega}_j$
should match $2\pi\Omega_{kk}^{-1}(\widetilde{\omega}_j)$, as demonstrated in \eqref{eq:cond_var_d}. We then define the discrepancy measure for each estimated chain component $\widehat{\tau}_g$ and any node set $\cM\subset[p]\backslash \widehat{\tau}_g$ as
\begin{equation}
    \label{eq:discrepancy}
    \widehat{\cD}(\widehat{\tau}_g,\cM)=\max_{k\in \widehat{\tau}_g} \max_{j\in[M]}\big|\widehat{f}_{x,kk}(\widetilde{\omega}_j)-\widehat{\bbf}_{x,k\cM}(\widetilde{\omega}_j)\widehat{\bbf}_{x,\cM\cM}^{-1}(\widetilde{\omega}_j)\widehat{\bbf}_{x,\cM k}(\widetilde{\omega}_j)-\widehat{\Omega}^{-1}_{kk}(\widetilde{\omega}_j)\big|,
\end{equation}
where $\widehat{f}_{x,kk}(\widetilde{\omega}_j) - \widehat{\bbf}_{x,k\cM}(\widetilde{\omega}_j)\widehat{\bbf}_{x,\cM\cM}^{-1}(\widetilde{\omega}_j)\widehat{\bbf}_{x,\cM k}(\widetilde{\omega}_j)$ and $\widehat{\Omega}^{-1}_{kk}(\widetilde{\omega}_j)$ are the estimated asymptotic conditional variances (divided by $2\pi$) of node $k\in\widehat{\tau}_g$ at frequency $\widetilde{\omega}_j$ given $\cM$ and given
its all parent nodes, respectively. 
One would expect that $\widehat{\cD}(\widehat{\tau}_g,\cM)$ to be close to $0$ for large $T$ if $\cM$ includes all parent chain components of $\widehat{\tau}_g$.  The iterative top-down ordering procedure then proceeds as follows.
We begin by computing $\widehat{\cD}(\widehat{\tau}_g,\emptyset)=\max_{k\in\widehat{\tau}_g}\max_{j\in[M]}\big|\widehat{f}_{x,kk}(\widetilde{\omega}_j)-\widehat{\Omega}^{-1}_{kk}(\widetilde{\omega}_j)\big|$ for each chain component $\widehat{\tau}_g$, and then select the first chain component by $\widehat{\pi}_1=\argmin_{g\in[\widehat{G}]} \widehat{\cD}(\widehat{\tau}_g, \emptyset)$. 
For each $s < \widehat{G}$, we define $\widehat{\cM}_s=\bigcup_{r=1}^{s}\widehat{\tau}_{\widehat{\pi}_r}$ and recursively select
$\widehat{\pi}_{s+1}=\argmin_{g\in[\widehat{G}]\backslash \bigcup_{r=1}^{s}\widehat{\pi}_r} \widehat{\cD}\big(\widehat{\tau}_g,\widehat{\cM}_s\big)$. Repeat this step for all $s$. We finally obtain the  estimated causal ordering $\widehat{\bpi}=(\widehat{\pi}_1, \dots, \widehat{\pi}_{\widehat{G}})$. 
Such procedure shares a similar spirit with
\cite{chen2019causal,zhao2024identifiability} in determining the causal ordering.

In the third stage, we estimate the coefficient matrices $\bA$ and $\bB$, and consequently the directed edge set $\cE_d$, based on the estimated causal ordering $\widehat{\bpi}$ of the chain components. Recall that 
$\widehat{\cM}_{g-1}$ consists of all the parent chain components of $\widehat{\tau}_{\widehat{\pi}_g}$. By construction, any directed edges to $\widehat{\tau}_{\widehat{\pi}_g}$ can 
only originate from nodes in $\widehat{\cM}_{g-1}$. Accordingly, we compute the intermediate estimators $\widehat{\bA}^{\reg}$ and $\widehat{\bB}^{\reg}$, whose sub-matrices $\widehat{\bA}_{\widehat{\tau}_{\hat{\pi}_g},\widehat{\cM}_{g-1}}^{\reg}$ and $\widehat{\bB}_{\widehat{\tau}_{\hat{\pi}_g}, \widehat{\cM}_{g-1}}^{\reg}$ are obtained via a multivariate regression of $\bx_{t,\widehat{\tau}_{\hat{\pi}_g}}$ on $\by_{t,\widehat{\cM}_{g-1}}:=\big(\bx_{t,\widehat{\cM}_{g-1}}^{\T},\bx_{t-1,\widehat{\cM}_{g-1}}^{\T}\big)^{\T}$ for $t=2,\dots,T$. Specifically, 
$$
\Big(\widehat{\bA}_{\widehat{\tau}_{\hat{\pi}_g},\widehat{\cM}_{g-1}}^{\reg},\widehat{\bB}_{\widehat{\tau}_{\hat{\pi}_g}, \widehat{\cM}_{g-1}}^{\reg}\Big)=
\bigg\{\sum_{t=2}^{T}\bx_{t,\widehat{\tau}_{\hat{\pi}_g}}\by_{t,\widehat{\cM}_{g-1}}^{\T}\bigg\}\cdot
\bigg\{\sum_{t=2}^{T}\by_{t,\widehat{\cM}_{g-1}}\by_{t,\widehat{\cM}_{g-1}}^{\T}\bigg\}^{-1}.
$$
The consistency of the intermediate estimators is ensured by the weak exogeneity condition in Assumption~\ref{ass:cond_mean}. 
We next apply the singular value hard thresholding on $\widehat{\bA}^{\reg}$.
Let $\widehat{\bA}^{\reg} = \widehat\bU^{\reg} \widehat\bD^{\reg} (\widehat\bV^{\reg})^{\T}$ be the singular value decomposition of $\widehat{\bA}^{\reg}$, 
where $\widehat\bD^{\reg}  = \mathrm{diag}(\hat d_1,\dots,\hat d_p)$. 
We  compute $\widehat{\bA}^{\svd}=\cS_{\kappa_T}^{\hard}(\widehat{\bA}^{\reg}) := \widehat\bU^{\reg} \widehat\bD^{\svd} (\widehat\bV^{\reg})^{\T},$ where $\widehat\bD^{\svd} =  \mathrm{diag}\{\hat d_1 \cdot I(\hat d_1>\kappa_T),\dots,\hat d_p \cdot I(\hat d_p>\kappa_T)\}$ for some pre-specified $\kappa_T>0$ and $I(\cdot)$ denotes the indicator function.
The final estimate $\widehat{\bA}=(\widehat{A}_{kl})_{p \times p}$ is obtained by applying elementwise hard thresholding to $\widehat{\bA}^{\svd}$ while preserving the estimated causal ordering, 
i.e., $\widehat{A}_{kl}=\widehat{A}_{kl}^{\svd}\cdot I(|\widehat{A}_{kl}^{\svd}|>\nu_T)\cdot I(k\in\widehat{\tau}_{\hat{\pi}_g},l\in\widehat{\tau}_{\hat{\pi}_s},g>s)$ for some pre-specified $\nu_T>0$. Applying the same procedure to $\widehat{\bB}^{\reg}$ yields the estimate  $\widehat{\bB}$. We then obtain the estimated directed edge set as $\widehat{\cE}_d=\big\{(l\to k)\in\cN\times\cN:\widehat{A}_{kl}\neq0\ {\rm or}\ \widehat{B}_{kl}\neq0\big\}.$  

We summarize the proposed three-stage learning algorithm in  Algorithm~\ref{alg:1} below.

\begin{algorithm}[ht]
\spacingset{1.6}
\caption{Learning time series Gaussian chain graph models}
\label{alg:1}
\begin{algorithmic}[1]
\State \textbf{Input}: Data $\{\bx_t\}_{t\in[T]}.$ 

\Statex \textbf{Stage 1: Estimate the undirected edges} \label{al.s1}
\State Transform $\{\bx_t\}_{t\in[T]}$ to $\{\bd_{x}(\omega_j)\}_{j\in[T]}$ via the DFT as in \eqref{eq:dft}.

\State Compute the averaged periodogram estimator $\widehat{\bbf}_x(\widetilde{\omega}_j)$ in \eqref{eq:spec_est}.


\State Solve the optimization problem in \eqref{eq:min} to obtain
the estimates $(\widehat{\bOmega},\widehat{\bL})$ by  Algorithm~\ref{alg:admm}.


\State Let $\widehat{\cE}_u=\big\{(l-k)\in\cN\times\cN:\exists j\in[M],\widehat{\Omega}_{kl}(\widetilde{\omega}_j)\neq0\big\}.$

\Statex \textbf{Stage 2: Determine the causal ordering}\label{al.s2}
\State Partition $\cN$ into estimated chain components $\{\widehat{\tau}_1, \dots, \widehat{\tau}_{\widehat{G}}\}$ based on $\widehat{\cE}_u$.


\State For $s\in[\widehat{G}]$, recursively select $\widehat{\pi}_s$ using the discrepancy measure in \eqref{eq:discrepancy}.

\State Determine the causal ordering as $\widehat{\bpi}=(\widehat{\pi}_1,\dots,\widehat{\pi}_{\widehat{G}}).$

\Statex \textbf{Stage 3: Estimate the directed edges} \label{al.s3}

\State Regress $\bx_{t,\widehat{\tau}_{\widehat{\pi}_g}}$ on $\big(\bx_{t,\widehat{\cM}_{g-1}}^{\T},\bx_{t-1,\widehat{\cM}_{g-1}}^{\T})^{\T}$ for $g\in[\widehat{G}]$ to calculate $\widehat{\bA}^{\reg}$ and $\widehat{\bB}^{\reg}$.

\State Truncate the small singular values to obtain $\widehat{\bA}^{\svd}=\cS_{\kappa_T}^{\hard}(\widehat{\bA}^{\reg})$ and $\widehat{\bB}^{\svd}=\cS_{\kappa_T}^{\hard}(\widehat{\bB}^{\reg})$.

\State Construct $\widehat{\bA}=(\widehat{A}_{kl})_{p\times p}$ and $\widehat{\bB}=(\widehat{B}_{kl})_{p\times p}$, where $\widehat{A}_{kl}=\widehat{A}_{kl}^{\svd}\cdot I(|\widehat{A}_{kl}^{\svd}|>\nu_T)\cdot I(k\in\widehat{\tau}_{\hat{\pi}_g},l\in\widehat{\tau}_{\hat{\pi}_s},g>s)$ and $\widehat{B}_{kl}=\widehat{B}_{kl}^{\svd}\cdot I(|\widehat{B}_{kl}^{\svd}|>\nu_T)\cdot I(k\in\widehat{\tau}_{\hat{\pi}_g},l\in\widehat{\tau}_{\hat{\pi}_s},g>s).$

\State Let $\widehat{\cE}_d=\big\{(l\to k)\in\cN\times\cN:\widehat{A}_{kl}\neq0\ {\rm or}\ \widehat{B}_{kl}\neq0\big\},$ and $\widehat{\cE}=\widehat{\cE}_u\bigcup \widehat{\cE}_d$.


\State \textbf{Output}: The estimated chain graph $\widehat{\cG}=(\cN,\widehat{\cE})$, the estimated chain components $\{\widehat{\tau}_1,\dots,\widehat{\tau}_{\widehat{G}}\}$ and the estimated causal ordering $\widehat{\bpi}$.
\end{algorithmic}
\end{algorithm}

\begin{remark}
\label{rm.CIG}
Given the estimates obtained from Algorithm~\ref{alg:1},
    we can further compute the residuals for each estimated chain component $\widehat{\tau}_{\hat{\pi}_g}$ via
$$\widehat{\be}_{t,\widehat{\tau}_{\hat{\pi}_g}}=\bx_{\widehat{\tau}_{\hat{\pi}_g}}-\widehat{\bA}_{\widehat{\tau}_{\hat{\pi}_g},\widehat{\cM}_{g-1}}\bx_{t,\widehat{\cM}_{g-1}}-\widehat{\bB}_{\widehat{\tau}_{\hat{\pi}_g},\widehat{\cM}_{g-1}}\bx_{t-1,\widehat{\cM}_{g-1}},\quad t=2,\dots,T.
    $$
    As discussed in Remark~\ref{rmk:e_var}, each $\widehat{\be}_{t,\widehat{\tau}_{\hat{\pi}_g}}$ can then be modeled to characterize the contemporaneous conditional dependencies and dynamic Granger‐causal relations within the corresponding chain component. For simplicity, we adopt a VAR(1) specification and,  for each $\widehat{\tau}_{\hat{\pi}_g}$, apply the algorithm of \cite{songsiri2010topology} to fit the model subject to the group sparsity structure of $\{\widehat \bOmega_{\widehat{\tau}_{\hat{\pi}_g}, \widehat{\tau}_{\hat{\pi}_g}}(\widetilde \omega_j): j \in [M]\}$. This yields the estimated coefficient matrix $\widehat{\bC}_{\widehat{\tau}_{\hat{\pi}_g},\widehat{\tau}_{\hat{\pi}_g}}$ and precision matrix $\widehat{\bSigma}_{\varepsilon,\widehat{\tau}_{\hat{\pi}_g},\widehat{\tau}_{\hat{\pi}_g}}^{-1}$, from which the associated Granger causality graph can be recovered.
\end{remark}


\subsection{ADMM algorithm}
\label{subsec:admm}

To ensure that the objective function of \eqref{eq:min} is separable, we rewrite $\ell_M(\widetilde{\bOmega}+\widetilde{\bL})$ as $\ell_M(\widetilde{\bTheta})$ and introduce the constraints $\widetilde{\bTheta}=\widetilde{\bOmega}+\widetilde{\bL}$. Following \cite{xue2012positive}, we replace the positive definite constraints on $\bOmega(\widetilde{\omega}_j)$ with slightly stronger constraints $\bOmega(\widetilde{\omega}_j)\succcurlyeq\varrho\bI_p$ for all $j\in[M]$ and some small $\varrho>0$, and reformulate  \eqref{eq:min} as follows:
\begin{equation}
    \label{eq:min_new}
    \begin{aligned}
        (\widehat{\bTheta},\widehat{\bOmega},\widehat{\bL})
        =&\argmin_{\widetilde{\bTheta},\widetilde{\bOmega},\widetilde{\bL}}
        -\ell_M(\widetilde{\bTheta})+P_1(\widetilde{\bOmega},\lambda_{1T})+P_2(\widetilde{\bL},\lambda_{2T}),\\
        &\text{s.t.}\quad\bTheta(\widetilde{\omega}_j)=\bOmega(\widetilde{\omega}_j) + \bL(\widetilde{\omega}_j),\quad\bTheta(\widetilde{\omega}_j)\succ\bzero,\quad\bOmega(\widetilde{\omega}_j)\succcurlyeq\varrho\bI_p,\quad j\in[M].
    \end{aligned}
\end{equation}
The augmented Lagrangian function of \eqref{eq:min_new} is then defined as 
$$
\begin{aligned}
    L_{\rho}(\widetilde{\bTheta},\widetilde{\bOmega},\widetilde{\bL},\cU):=&-\ell_M(\widetilde{\bTheta})
    +P_1(\widetilde{\bOmega},\lambda_{1T})+P_2(\widetilde{\bL},\lambda_{2T})
    +\sum_{j=1}^{M}\tr[\bU(\widetilde{\omega}_j)^{\H}\{\bTheta(\widetilde{\omega}_j)-\bOmega(\widetilde{\omega}_j)-\bL(\widetilde{\omega}_j)\}]\\
    &+\frac{\rho}{2}\sum_{j=1}^{M}\|\bTheta(\widetilde{\omega}_j)-\bOmega(\widetilde{\omega}_j)-\bL(\widetilde{\omega}_j)\|_{\F}^2,
\end{aligned}
$$
where $\cU\in\eC^{p\times p\times M}$ is a complex-valued tensor of the dual variable with $\cU_{::j}=\bU(\widetilde{\omega}_j)$ for $j\in[M]$, and $\rho>0$ is a penalty parameter. The corresponding dual problem is 
\begin{equation*}
    \max_{\cU}\min_{\bTheta(\widetilde{\omega}_j)\succ \bzero,\bOmega(\widetilde{\omega}_j)\succcurlyeq \varrho\bI_p,\widetilde{\bL}}L_{\rho}(\widetilde{\bTheta},\widetilde{\bOmega},\widetilde{\bL},\cU),
\end{equation*}
which can be solved efficiently by an ADMM algorithm, as outlined in Algorithm~\ref{alg:admm}. In particular, each iteration of the algorithm requires optimizing scalar objective functions of complex-valued tensors and thus the update rules are derived with the aid of Wirtinger calculus and the associated Wirtinger subdifferential \cite[]{schreier2010statistical}; see Section~{\color{blue}B.1} of the supplementary material for further discussion. For brevity, the detailed update rules for each ADMM step are presented in Sections~{\color{blue}B.2}--{\color{blue}B.4} of the supplementary material.

\begin{center}
	\begin{algorithm}[!t] \caption{\label{alg:admm}{ADMM algorithm to solve \eqref{eq:min}}}
		\begin{enumerate}
			\item[\small 1:] \textbf{Input:} Initial estimators $\widetilde{\bOmega}^{(0)}$ of $\widetilde{\bOmega}$ and $\widetilde{\bL}^{(0)}$ of $\widetilde{\bL}$, $\cU^{(0)}=\bzero.$ 
			\item[\small 2:] For $u=0,1,...$ do 	
			\begin{itemize}		
				\item[] (a)~ $\widetilde{\bTheta}^{(u+1)}=\argmin_{\bTheta(\widetilde{\omega}_j)\succ \bzero}L_{\rho}(\widetilde{\bTheta},\widetilde{\bOmega}^{(u)},\widetilde{\bL}^{(u)},\cU^{(u)})$ as in Section~{\color{blue}B.2}.
				\item[] (b)~ $\widetilde{\bOmega}^{(u+1)}:=\argmin_{\bOmega(\widetilde{\omega}_j)\succcurlyeq\varrho\bI_p}L_{\rho}(\widetilde{\bTheta}^{(u+1)},\widetilde{\bOmega},\widetilde{\bL}^{(u)},\cU^{(u)})$ as in Section~{\color{blue}B.3}.
				\item[] (c)~  $\widetilde{\bL}^{(u+1)}:=\argmin_{\widetilde{\bL}}L_{\rho}(\widetilde{\bTheta}^{(u+1)},\widetilde{\bOmega}^{(u+1)},\widetilde{\bL},\cU^{(u)})$ as in Section~{\color{blue}B.4}.
				\item[] (d)~  $\cU^{(u+1)}:=\cU^{u}+\rho(\widetilde{\bTheta}^{(u+1)}-\widetilde{\bOmega}^{(u+1)}-\widetilde{\bL}^{(u+1)}).$
			\end{itemize}
			\item[] end do until convergence.
			\item[\small 3:] \textbf{Output:} The converged estimates $(\widehat{\bOmega},\widehat{\bL})$. 
		\end{enumerate}
	\end{algorithm}		
\end{center}

\vspace{-1.2em}

\section{Asymptotic properties}
\label{sec:theory}
This section presents the theoretical analysis of our proposed three-stage estimation procedure. 
We start by imposing some regularity assumptions. 
Let $\lambda_{\min}(\cdot)$ denote the smallest eigenvalue of a symmetric or Hermitian matrix.

\begin{assumption}
    \label{ass:spectral}
    There exist constants $\alpha>1$ and $\delta_1,\delta_2>0$ such that: (i) $\sum_{h\in\eZ}|h|^{\alpha}\|\bSigma_x(h)\|<\infty$; (ii) $\delta_1<\lambda_{\min}\big(\bbf_x(\omega)\big)\le\lambda_{\max}\big(\bbf_x(\omega)\big)<\delta_2$ for all $\omega\in(0,2\pi].$
\end{assumption}

Assumption~\ref{ass:spectral}(i) is standard in the Whittle likelihood literature \citep[e.g.,][]{choudhuri2004contiguity,kirch2019beyond}, where $\alpha$ characterizes the strength of temporal dependence in $\{\bx_t\}$. 
This condition can be relaxed by requiring $\alpha>0$, which leads to slower convergence rates for $\alpha\in (0,1)$ compared with those established in Theorem~\ref{thm:undir}.
Moreover, Assumption~\ref{ass:spectral}(i) implies the standard absolutely summable condition $\sum_{h\in\eZ}\|\bSigma_x(h)\|<\infty$ and thus $\bbf_x(\cdot)\in\cC((0,2\pi];\cH_+^p)$. 
Assumption~\ref{ass:spectral}(ii) ensures that $\bbf_x^{-1}(\omega)$ exists for all $\omega\in(0,2\pi]$, $\bbf_x^{-1}(\cdot)\in\cC((0,2\pi];\cH_{++}^p)$, and both $\|\bbf_x(\omega)\|$ and $\|\bbf_x^{-1}(\omega)\|$ are uniformly bounded. See \cite{jung2015graphical,tugnait2022sparse} for similar assumptions in  time series graphical models. Since $\be_t=(\bI_p-\bA)\bx_t-\bB\bx_{t-1}$, Assumption~\ref{ass:spectral} naturally applies to $\{\be_t\}$. 

Before imposing the condition required to establish the selection consistency of $\widehat{\bOmega}$ and the rank consistency of $\widehat{\bL}$ in \eqref{eq:min}, we first introduce  some definitions.  
Define $\bar{\ell}(\bTheta):=(2\pi)^{-1}\int_{0}^{2\pi}[\log \det\{\bTheta(\omega)\} -\tr\{\bTheta(\omega)\bbf_x(\omega)\}]\domega$ as a functional of the matrix-valued function $\bTheta(\cdot)$. We also write $\bar{\ell}(\bTheta)=\bar{\ell}(\bOmega+\bL):=\bar{\ell}(\bOmega,\bL)$. By Lemma~{\color{blue}B.6} of the supplementary material, $(\bOmega_0,\bL_0)$ is the unique maximizer of $\bar{\ell}(\bOmega,\bL)$ within its localization set. Then, we define the following bilinear matrix-valued operator $\cI_0(\cdot,\cdot):\cC((0,2\pi];\cH^p)\times\cC((0,2\pi];\cH^p)\to\eR$ satisfying
$
\cI_0(\bDelta,\bDelta'):=-\nabla^2\bar{\ell}(\bTheta_0)(\bDelta,\bDelta'), 
$
where $\nabla^2\bar{\ell}(\bTheta_0)(\cdot,\cdot)$ denotes the second-order derivative of $\bar{\ell}(\bTheta)$ at $\bTheta=\bTheta_0$ \cite[]{higham2008functions}, 
 and $\vvec(\cdot)$ denotes the vectorization operator. 
The operator $\cI_0(\cdot,\cdot)$ satisfies the pointwise representation $\big(\cI_0\vvec(\bDelta)\big)(\omega)=\vvec^{-1}\big[\{\bTheta_0^{-1}(\omega)\otimes\bTheta_0^{-1}(\omega)\}\vvec\{\bDelta(\omega)\}\big]\in\cH^p$ for $\omega\in(0,2\pi]$, where $\vvec^{-1}(\cdot)$ denotes the inverse vectorization operator. For a linear subspace $\cC_1\subset\cC((0,2\pi];\cH^p),$ define its orthogonal complement as  $\cC_1^{\perp}:=\big\{\bOmega'(\cdot)\in\cC((0,2\pi];\cH^p):\int_{0}^{2\pi}\tr\{\bOmega'(\omega)^{\H}\bOmega(\omega)\}=0,\forall\bOmega(\cdot)\in\cC_1\big\}$, 
and $\cP_{\cC_1}(\cdot)$ denote the projection operator onto $\cC_1.$ We further define two linear operators $F:\cS(\bOmega_0)\times\cT(\bL_0)\to\cS(\bOmega_0)\times\cT(\bL_0)$ and $F^{\perp}:\cS(\bOmega_0)\times\cT(\bL_0)\to\cS(\bOmega_0)^{\perp}\times\cT(\bL_0)^{\perp}$ such that
$$
\begin{aligned}
    &F(\bOmega,\bL):=\left(\cP_{\cS(\bOmega_0)}\{\cI_0\vvec(\bOmega+\bL)\},\cP_{\cT(\bL_0)}\{\cI_0\vvec(\bOmega+\bL)\}\right),\\
    &F^{\perp}(\bOmega,\bL):=\left(\cP_{\cS(\bOmega_0)^{\perp}}\{\cI_0\vvec(\bOmega+\bL)\},\cP_{\cT(\bL_0)^{\perp}}\{\cI_0\vvec(\bOmega+\bL)\}\right).
\end{aligned}
$$
Lemma~{\color{blue}S.5} of the supplementary material 
guarantees that $F$ is invertible, and thus $F^{-1}$ is well defined. 
Based on the dual norms of the penalty terms $P_1$ and $P_2$ in \eqref{eq:min}, for any $\bOmega,\bL\in\cC((0,2\pi];\cH^p)$, we define $g_{\gamma}(\bOmega,\bL):=\max\{\vvvert\bOmega\vvvert_{\max},\vvvert\bL\vvvert/\gamma\}$ for some positive constant $\gamma$, where $\vvvert\bL\vvvert :=\sup_{\omega\in(0,2\pi]}\|\bL(\omega)\|$. 
Moreover, define $\Phi(\bOmega)\in\cC((0,2\pi];\cH^p)$ such that $\big(\Phi(\bOmega)\big)_{kl}(\cdot)=\Omega_{kl}(\cdot)/\big\{(2\pi)^{-1}\int_{0}^{2\pi}|\Omega_{kl}(\omega)|^2\domega\big\}^{1/2}$ if $(k,l)\in\gsupp(\bOmega)\cap\{k\neq l\}$ and 0 otherwise, and $\Psi(\bD)\in\cC((0,2\pi];\cH^R)$ such that $\big(\Psi(\bD)\big)(\cdot)=\diag\Big(D_{11}(\cdot)/\big\{(2\pi)^{-1}\int_{0}^{2\pi}|D_{11}(\omega)|^2\domega\big\},$ $\dots,D_{RR}(\cdot)/\big\{(2\pi)^{-1}\int_{0}^{2\pi}|D_{RR}(\omega)|^2\domega\big\}\Big)$.
Let $\bL_0(\cdot)=\bU_0\bD_0(\cdot)\bU_0^{\H}$ be the eigen-decomposition of $\bL_0(\cdot)$ with $\bU_0\in\eC^{p\times R_0}$. 

\begin{assumption}
    \label{ass:irrepresentable}
    $g_{\gamma}\big(F^{\perp}F^{-1}(\Phi(\bOmega_0),\gamma\bU_0\Psi(\bD_0)\bU_0^{\H})\big)<1$ for some constant $\gamma>0.$
\end{assumption}

Assumption~\ref{ass:irrepresentable} is a new irrepresentable condition, 
which plays a key role in establishing the selection and rank consistency of $(\widehat{\bOmega},\widehat{\bL})$ through the group lasso and tensor-unfolding nuclear norm penalties, via the primal-dual witness technique in our proof. 
See also \cite{chandrasekaran2012latent,zhao2024identifiability} for similar conditions.
To aid intuition, 
we consider the special case, where $\bTheta_0(\omega)=\bI_p$ for $\omega\in(0,2\pi]$ 
and the subspace $\cS(\bOmega_0)$ is orthogonal to $\cT(\bL_0)$. Assumption~\ref{ass:irrepresentable} then reduces to $\max\{\gamma\vvvert\bU_0\Psi(\bD_0)\bU_0^{\H}\vvvert_{\max},\vvvert\Phi(\bOmega_0)\vvvert/\gamma\}<1,$ which indicates that $\bU_0$ is not very sparse since $\sum_{i=1}^{p}U_{ij}^2=1$ for each $j$, and that  $\Phi(\bOmega_0)$ is not low-rank since $\|\Phi(\bOmega_0)(\omega)\|\ge\|\Phi(\bOmega_0)(\omega)\|_{\F}/\rank\big(\Phi(\bOmega_0)(\omega)\big)$. 
It is noteworthy that the continuous functions $\bOmega(\cdot)$ and $\bL(\cdot)$ in the optimization problem \eqref{eq:min} are evaluated only at discrete Fourier frequencies $\widetilde{\omega}_j$ for $j\in[M]$ with $M\to\infty$. By exploiting the Riemann sum approximations and the Lipschitz continuity of the operators $F^{\perp}$ and $F^{-1}$, we show that the discretized counterpart of our irrepresentable condition holds asymptotically; see Lemma~{\color{blue}S.14} of the supplementary material.

Let $\cE_{u,0},\cE_{d,0}$ and $\cG_0$ denote the true values of $\cE_u,\cE_d$ and $\cG$, respectively. Let $\bPi_0$ be the set of all possible true causal orderings. We are now ready to present the main theorems.

\begin{theorem}
    \label{thm:undir}
    Suppose that the assumptions of Theorem \ref{thm:iden} and Assumptions~\ref{ass:spectral}--\ref{ass:irrepresentable} hold. Let $m\asymp (\log T)^{1/3}T^{2/3}$, $\lambda_{1T}\asymp T^{-1/3+\eta}$ for a sufficiently small constant $\eta>0$, and $\lambda_{2T}=\gamma\lambda_{1T}$ with $\gamma$ specified in Assumption~\ref{ass:irrepresentable}. Then, with probability tending to one, \eqref{eq:min} has a unique solution $(\widehat{\bOmega},\widehat{\bL}).$ Letting $\widehat{\bOmega}(\widetilde{\omega}_j)=\widehat{\bOmega}_{::j}$ and $\widehat{\bL}(\widetilde{\omega}_j)=\widehat{\bL}_{::j}$ for $j\in[M]$, we have:\\
    (i) $\max_{j\in[M]}\|\widehat{\bOmega}(\widetilde{\omega}_j)-\bOmega_0(\widetilde{\omega}_j)\|_{\max}=O_p(T^{-1/3+2\eta})$ and $\eP\big\{\gsupp_M(\widehat{\bOmega})=\gsupp(\bOmega_0)\big\}\to1$ as $T\to\infty$, where $\gsupp_M(\widehat{\bOmega}):=\{(k,l):\exists j\in[M],\widehat{\Omega}_{kl}(\widetilde{\omega}_j)\neq 0\}$; \\
    (ii) $\max_{j\in[M]}\|\widehat{\bL}(\widetilde{\omega}_j)-\bL_0(\widetilde{\omega}_j)\|_{\max}=O_p(T^{-1/3+2\eta})$ and $\eP\Big[\bigcap_{j=1}^{M}\big\{\rank\big(\widehat{\bL}(\widetilde{\omega}_j)\big)=\rank\big(\bL_0(\widetilde{\omega}_j)\big)\big\}\Big]\to1$ as $T\to\infty;$ \\
    (iii) $\eP(\widehat{\cE}_u=\cE_{u,0})\to1$ as $T\to\infty$.
\end{theorem}

Theorem~\ref{thm:undir} shows that $\widehat{\bOmega}$ achieves both estimation and selection consistency, while $\widehat{\bL}$ exhibits both estimation and rank consistency. Consequently, the undirected edge set $\cE_{u,0}$ and the group low-rank structure of $\bL_0(\cdot)$ can be recovered exactly with probability tending to one. The half-block size $m$ is selected to balance the bias-variance tradeoff, thereby yielding the fastest uniform convergence rate of the averaged periodogram estimator $\widehat{\bbf}_x(\widetilde{\omega}_j)$ over $j \in [M]$; see Remark~{\color{blue}S.2} of the supplementary material. 
Compared with the independent setting in \cite{zhao2024identifiability}, the temporal dependence introduces additional complexities in the theoretical analysis. Specifically, 
the nonparametric spectral density estimator $\widehat{\bbf}_x(\widetilde{\omega}_j)$ converges more slowly than the sample covariance matrix used in the independent case, which in turn leads to slower convergence rates of $(\widehat{\bOmega},\widehat{\bL})$.

\begin{theorem}
    \label{thm:order}
    Suppose that the assumptions of Theorem \ref{thm:undir} hold. Then, we have $\eP(\widehat{\bpi}\in\bPi_0)\to1$ as $T\to\infty.$
\end{theorem}   

Supported by Theorem~\ref{thm:order}, we assume that the estimated causal ordering $\widehat{\bpi}=\bpi_0\in\bPi_0$ with high probability, and that $\bA_0$ and $\bB_0$ are expressed under this true causal ordering $\bpi_0$. 

\begin{theorem}
    \label{thm:dir}
    Suppose that the assumptions of Theorem \ref{thm:order} hold. Let $\kappa_T\asymp T^{-1/2}$ and $\nu_T\asymp T^{-1/2+\zeta}$ with a positive constant $\zeta<1/2$. Then, we have:\\
    (i) $\|\widehat{\bA}-\bA_0\|_{\max}=O_p(T^{-1/2})$ and $\eP\{\sign(\widehat{\bA})=\sign(\bA_0)\}\to1$ as $T\to\infty;$\\
    (ii) $\|\widehat{\bB}-\bB_0\|_{\max}=O_p(T^{-1/2})$ and $\eP\{\sign(\widehat{\bB})=\sign(\bB_0)\}\to1$ as $T\to\infty;$\\
    (iii) $\eP(\widehat{\cE}_d=\cE_{d,0})\to1$ and $\eP(\widehat{\cG}=\cG_0)\to1$ as $T\to\infty$.
\end{theorem}

Theorem~\ref{thm:dir} establishes the estimation and sign consistency of both $\widehat{\bA}$ and $\widehat{\bB}$, which leads to the exact recovery of the directed edge set $\cE_{d,0}$ and, together with Theorem~\ref{thm:undir}, the full time series chain graph $\cG_0$ with high probability. Notably, both $\widehat{\bA}$ and $\widehat{\bB}$ achieve the parametric $\sqrt{T}$ rate, which is faster than the rate in Theorem 3 of \cite{zhao2024identifiability}. This improvement arises from our use of a tighter error bound for the sample (auto)covariance matrices.

\section{Simulation studies}
\label{sec:sim}
We conduct a series of simulations to evaluate the performance of our proposed TSCG method.  Specifically, we consider model~\eqref{eq:model} under two designs of the time series chain graph. 
The undirected and directed edges in $\cG$ are generated as follows.

\newcounter{bean}
\setcounter{bean}{0}
\begin{center}
	\begin{list}
		{\textsc{Design} \arabic{bean}}{\usecounter{bean}}
  
		\item \label{case1}(two-layer).  We first split the 
$p$ dimensions into two layers, $\cL_1 = \{1, \dots, \lceil 0.1p\rceil\}$ and $\cL_2 = \{\lceil 0.1p\rceil+1, \dots, p\}$. Within each layer, we connect each pair of nodes by an undirected edge with probability 0.02. 
Directed edges are then added from nodes in $\cL_1$ to nodes in $\cL_2$ with probability 0.8.
		\item \label{case2} (random-order). We first connect each pair of nodes by an undirected edge with probability \(0.02\), and let \(\{\tau_1,\dots,\tau_G\}\) denote the resulting chain components. We then adopt this as the causal order, i.e., \((\pi_1,\dots,\pi_G)=(1,\dots,G)\), and allow directed edges only from \(\tau_g\) to \(\tau_h\) for \(h>g\).
Within each component \(\tau_g\), nodes are independently selected as hubs with probability \(0.1\).
For each hub \(l\in\tau_g\) and each node \(k\in \bigcup_{h=g+1}^{G}\tau_h\), we include the directed edge \(l\to k\) with probability \(0.8\).
	\end{list}
\end{center} 
For each $g\in[G]$, we then generate the component process $\{\be_{t,\tau_g}\}_{t \in [T]}$ from a VAR(1) model, i.e.,
$\be_{t, \tau_g} = \bC_{\tau_g}\be_{t-1, \tau_g} + \bvar_{t,\tau_g}$, where $\bvarepsilon_{t,\tau_g} \sim \mathcal{N}_r(\bzero, \bSigma_{\varepsilon,\tau_g,\tau_g})$ and $\bC_{\tau_g} \in \mathbb{R}^{|\tau_g| \times |\tau_g|}$.
To ensure stationarity, we set $\bC_{\tau_g} = \iota\widecheck \bC_{\tau_g}/\rho(\widecheck \bC_{\tau_g})$, where $\iota \sim$ $\text{Uniform}[0.5,1],$ $\rho(\widecheck \bC_{\tau_g})$ denotes the spectral radius of $\widecheck \bC_{\tau_g}$, and the entries of $\widecheck \bC_{\tau_g}$ are uniformly sampled from $[-1,-0.5] \cup [0.5,1]$.  Take $\bSigma_{\varepsilon,\tau_g,\tau_g} = (\bI_{|\tau_g|} - \bC_{\tau_g})(\bI_{|\tau_g|} - \bC_{\tau_g}^{\T})$. Hence,
$\bOmega_{\tau_g,\tau_g}(\omega)
=
2\pi
(\bI_{|\tau_g|}-\bC_{\tau_g}^{\T}e^{i\omega})
\bSigma_{\varepsilon,\tau_g,\tau_g}^{-1}
(\bI_{|\tau_g|}-\bC_{\tau_g}e^{-i\omega})$. 
The chain components of $\{\bx_t\}$, as encoded by $\{\bOmega(\omega):\omega\in(0,2\pi]\}$, then coincide with $\{\tau_g\}_{g=1}^G$, as each $\{\bx_{t,\tau_g}\}$ is verified to form a fully connected CIG for $g \in [G]$.
Based on the directed edge set, the nonzero entries of $\bA$ and $\bB$  are uniformly sampled from $[-1.5,-0.5] \cup [0.5,1.5]$.


We generate $T\in \{500,1000\}$ observations with $p \in \{30,60\}$ for each design and replicate each simulation 100 times. For implementation, we set $\eta=1/16$ and choose $m, \lambda_{1T},\lambda_{2T},\kappa_T,\nu_T$ proportional to the theoretical rates as suggested in {Theorems~\ref{thm:undir} and \ref{thm:dir}}. 
To assess the performance of the proposed TSCG learning method, we compute the averages of recall, precision and Matthews correlation coefficient (MCC) for the estimated undirected edge set $\widehat{\cE}_u$ and for the estimated directed edges corresponding to $\widehat{\bA}$ and $\widehat{\bB}$, respectively. We further examine the overall time series chain graph recovery using the structural Hamming
distance (SHD) \citep{tsamardinos2006}, defined as the minimum number of edge insertions, deletions, or orientation changes required to transform $\widehat \cG$ into the true $\cG$.
 Tables~\ref{err.table1} and \ref{err.table2} report the numerical summaries for Designs~\ref{case1} and \ref{case2}, respectively.
For comparison, we also implement the independent data chain graph learning method (ICG) of  \cite{zhao2024identifiability} 
using the R package \texttt{LearnCG} with its recommended tuning parameters. We subsequently estimate $\bA$ and $\bB$ as in Step 3 of Algorithm \ref{alg:1} with the causal ordering obtained from ICG.

Several conclusions can be drawn from Tables~\ref{err.table1} and \ref{err.table2}. First, TSCG consistently achieves high recall, precision, MCC and relatively small SHD across all settings, and its performance further improves as $T$ increases. This highlights the effectiveness of our method in accurately recovering the chain graph structure for time series data. 
Second, ICG performs poorly in identifying the undirected edge set, as indicated by the low MCC of $\widehat{\cE}_u$, which in turn leads to overall less accurate support recovery for $\bA$ and $\bB$. Recall that we take $\bSigma_{\varepsilon,\tau_g,\tau_g} = (\bI_{|\tau_g|} - \bC_{\tau_g})(\bI_{|\tau_g|} - \bC_{\tau_g}^{\T})$, which implies that $\bSigma_{e,\tau_g,\tau_g} = \bI_{|\tau_g|}$. 
The ICG method, originally developed for independent data and aimed at estimating $\bSigma_{e}^{-1}(0)$, fails to account for dynamic conditional dependencies and therefore cannot fully capture the CIGs in time series data.
Interestingly, ICG may occasionally yield reasonable support recovery for $\bA$ and $\bB$, even though its estimation of the undirected edge set remains unsatisfactory, as observed, e.g., when $p=60$ in Table~\ref{err.table1}.
This typically occurs when the recall of $\widehat{\cE}_u$ is close to zero while the precision is close to one, suggesting that ICG tends to split true chain components into multiple smaller sub-chain-components. When the resulting causal ordering still places nodes in $\tau_g$ ahead of those in $\tau_h$ for $h>g$, this over-segmentation does not necessarily worsen the estimation of $\bA$ and $\bB$. However, the uniformly high SHD for ICG across all settings reaffirms the inherent limitations of this covariance-based method when applied to time series data.

\begin{table}[tbp]

	\caption{\label{err.table1} The average (standard error) of recall, precision, MCC, and SHD across 100 simulation runs for Design~1.}
	\begin{center}
		\vspace{-0.5cm}
		\resizebox{6.6in}{!}{
\begin{tabular}{cccccccccccccccc}
\hline
\multirow{2}{*}{$(p,T)$} & \multirow{2}{*}{Method} &                      & \multicolumn{3}{c}{$\widehat{\cE}_u$} &  & \multicolumn{3}{c}{$\widehat{\bA}$} &  & \multicolumn{3}{c}{$\widehat{\bB}$} &  & \multirow{2}{*}{SHD} \\ \cline{4-6} \cline{8-10} \cline{12-14}
                         &                         &                      & Recall     & Precision    & MCC        &  & Recall    & Precision   & MCC       &  & Recall    & Precision   & MCC       &  &                      \\ \hline
$(30,500)$               & TSCG                    &                      & 0.748      & 0.933        & 0.829      &  & 0.685     & 0.955       & 0.796     &  & 0.694     & 0.955       & 0.801     &  & 20.170                \\
                         &                         &                      & (0.008)    & (0.007)      & (0.006)    &  & (0.008)   & (0.003)     & (0.006)   &  & (0.012)   & (0.004)     & (0.009)   &  & (0.790)               \\
                         & ICG                     &                      & 0.157      & 0.430         & 0.244      &  & 0.363     & 0.909       & 0.554     &  & 0.321     & 0.899       & 0.516     &  & 58.890                \\
\multicolumn{1}{l}{}     & \multicolumn{1}{l}{}    & \multicolumn{1}{l}{} & (0.006)    & (0.014)      & (0.009)    &  & (0.010)    & (0.006)     & (0.009)   &  & (0.010)    & (0.005)     & (0.009)   &  & (0.908)              \\
$(30,1000)$              & TSCG                    &                      & 0.858      & 0.971        & 0.909      &  & 0.786     & 0.963       & 0.860      &  & 0.801     & 0.980        & 0.878     &  & 12.930                \\
                         &                         &                      & (0.006)    & (0.004)      & (0.004)    &  & (0.004)   & (0.003)     & (0.003)   &  & (0.005)   & (0.001)     & (0.003)   &  & (0.254)              \\
                         & ICG                     &                      & 0.189      & 0.424        & 0.267      &  & 0.301     & 0.818       & 0.473     &  & 0.293     & 0.882       & 0.490      &  & 67.120                \\
                         & \multicolumn{1}{l}{}    & \multicolumn{1}{l}{} & (0.006)    & (0.010)       & (0.007)    &  & (0.008)   & (0.006)     & (0.008)   &  & (0.006)   & (0.005)     & (0.006)   &  & (0.678)              \\
$(60,500)$               & TSCG                    &                      & 0.453      & 0.891        & 0.628      &  & 0.701     & 0.928       & 0.795     &  & 0.692     & 0.905       & 0.778     &  & 89.030                \\
                         &                         &                      & (0.003)    & (0.004)      & (0.003)    &  & (0.003)   & (0.002)     & (0.002)   &  & (0.003)   & (0.002)     & (0.002)   &  & (0.736)              \\
                         & ICG                     &                      & 0.024      & 0.978        & 0.147      &  & 0.632     & 0.894       & 0.737     &  & 0.634     & 0.922       & 0.751     &  & 124.940               \\
                         & \multicolumn{1}{l}{}    & \multicolumn{1}{l}{} & (0.001)    & (0.010)       & (0.003)    &  & (0.002)   & (0.002)     & (0.002)   &  & (0.003)   & (0.001)     & (0.002)   &  & (0.647)              \\
$(60,1000)$              & TSCG                    &                      & 0.484      & 0.961        & 0.676      &  & 0.768     & 0.947       & 0.843     &  & 0.755     & 0.935       & 0.830      &  & 72.330                \\
                         &                         &                      & (0.003)    & (0.002)      & (0.002)    &  & (0.002)   & (0.001)     & (0.001)   &  & (0.002)   & (0.001)     & (0.002)   &  & (0.583)              \\
                         & ICG                     &                      & 0.024      & 0.973        & 0.147      &  & 0.741     & 0.890        & 0.800       &  & 0.741     & 0.935       & 0.821     &  & 101.380               \\
                         & \multicolumn{1}{l}{}    & \multicolumn{1}{l}{} & (0.001)    & (0.011)      & (0.003)    &  & (0.002)   & (0.001)     & (0.001)   &  & (0.002)   & (0.001)     & (0.001)   &  & (0.486)              \\ \hline         
\end{tabular}
		}	
	\end{center}
\end{table}

\begin{table}[tbp]

	\caption{\label{err.table2}The average (standard error) of recall, precision, MCC, and SHD across 100 simulation runs for Design~2.}
	\begin{center}
		\vspace{-0.5cm}
		\resizebox{6.6in}{!}{
\begin{tabular}{cccccccccccccccc}
\hline
\multirow{2}{*}{$(p,T)$} & \multirow{2}{*}{Method} &  & \multicolumn{3}{c}{$\widehat{\cE}_u$} &  & \multicolumn{3}{c}{$\widehat{\bA}$} &  & \multicolumn{3}{c}{$\widehat{\bB}$} &  & \multirow{2}{*}{SHD} \\ \cline{4-6} \cline{8-10} \cline{12-14}
                         &                         &  & Recall     & Precision    & MCC        &  & Recall    & Precision   & MCC       &  & Recall    & Precision   & MCC       &  &                      \\ \hline
$(30,500)$               & TSCG                    &  & 0.772      & 0.990         & 0.871      &  & 0.584     & 0.881       & 0.708     &  & 0.662     & 0.902       & 0.764     &  & 13.390                \\
                         &                         &  & (0.009)    & (0.004)      & (0.006)    &  & (0.006)   & (0.005)     & (0.005)   &  & (0.007)   & (0.005)     & (0.005)   &  & (0.242)              \\
                         & ICG                     &  & 0.343      & 0.456        & 0.385      &  & 0.384     & 0.907       & 0.556     &  & 0.439     & 0.883       & 0.578     &  & 25.700                 \\
                         &                         &  & (0.004)    & (0.012)      & (0.006)    &  & (0.019)   & (0.016)     & (0.020)    &  & (0.024)   & (0.019)     & (0.025)   &  & (1.192)              \\
$(30,1000)$              & TSCG                    &  & 0.833      & 0.997        & 0.910       &  & 0.640      & 0.927       & 0.762     &  & 0.726     & 0.937       & 0.818     &  & 11.210                \\
                         &                         &  & (0.001)    & (0.002)      & (0.001)    &  & (0.006)   & (0.003)     & (0.004)   &  & (0.006)   & (0.004)     & (0.004)   &  & (0.216)              \\
                         & ICG                     &  & 0.335      & 0.348        & 0.331      &  & 0.166     & 0.629       & 0.277     &  & 0.165     & 0.822       & 0.265     &  & 41.380                \\
                         &                         &  & (0.002)    & (0.006)      & (0.003)    &  & (0.020)    & (0.034)     & (0.024)   &  & (0.024)   & (0.032)     & (0.028)   &  & (1.101)              \\
$(60,500)$               & TSCG                    &  & 0.709      & 0.752        & 0.721      &  & 0.483     & 0.844       & 0.634     &  & 0.559     & 0.907       & 0.704     &  & 42.690                \\
                         &                         &  & (0.004)    & (0.004)      & (0.004)    &  & (0.011)   & (0.014)     & (0.012)   &  & (0.012)   & (0.012)     & (0.012)   &  & (0.627)              \\
                         & ICG                     &  & 0.137      & 0.583        & 0.272      &  & 0.302     & 0.926       & 0.523     &  & 0.312     & 0.984       & 0.549     &  & 77.700                 \\
                         &                         &  & (0.001)    & (0.002)      & (0.001)    &  & (0.008)   & (0.008)     & (0.008)   &  & (0.007)   & (0.004)     & (0.006)   &  & (0.253)              \\
$(60,1000)$              & TSCG                    &  & 0.776      & 0.828        & 0.794      &  & 0.493     & 0.902       & 0.661     &  & 0.537     & 0.968       & 0.713     &  & 33.100                 \\
                         &                         &  & (0.004)    & (0.002)      & (0.003)    &  & (0.013)   & (0.013)     & (0.013)   &  & (0.013)   & (0.011)     & (0.012)   &  & (0.547)              \\
                         & ICG                     &  & 0.140       & 0.575        & 0.272      &  & 0.308     & 0.924       & 0.528     &  & 0.317     & 0.995       & 0.557     &  & 78.190                \\
                         &                         &  & (0.001)    & (0.002)      & (0.001)    &  & (0.007)   & (0.009)     & (0.007)   &  & (0.006)   & (0.002)     & (0.005)   &  & (0.208)          \\   \hline
\end{tabular}
		}	
	\end{center}
\end{table}

\section{Real data analysis}
\label{sec:real}
In this section, we apply the proposed TSCG method to explore the relationships among U.S. macroeconomic and financial variables. The FRED-MD data (\url{https://www.stlouisfed.org/research/economists/mccracken/fred-databases}) contains eight groups of U.S.  economic indicators. To study the transmission of monetary policy, we focus on $p=66$ monthly time series from four groups: Housing (G1), Interest \& Exchange Rates (G2), Prices (G3), and Money \& Credit (G4), over the period June 2009 to May 2019 ($T=120$), prior to the COVID-19 pandemic.
The full list of variable codes and descriptions is provided in Table~{\color{blue}S.1} of the supplementary material. Following \cite{McCracken2016}, all series are transformed to be stationary and standardized before analysis.

Figure~\ref{fig:real.CIG} displays the estimated CIGs, where different colors denote the predefined groups. Notably, all undirected edges are detected within groups. In G1, the new private housing permits series in the Northeast (PERMITNE) is connected to housing starts in the same region (HOUSTNE), reflecting conditional dependencies in regional construction activity.  In G2, undirected edges appear among the 5- and 10-year Treasury yields (GS5, GS10) and Moody’s Aaa and Baa corporate bond yields (AAA, BAA), which implies the conditional dependencies among medium- and long-term government and corporate borrowing costs. We also find an edge connecting the 5- and 10-year term spreads over the federal funds rate (T5YFFM, T10YFFM).
Moreover, the trade-weighted U.S. dollar index (TWEXMMTH)
is connected to exchange rates against the Swiss franc (EXSZUSx), the British pound (EXUSUKx), and the Canadian dollar (EXCAUSx). See Section~{\color{blue}C} of the supplementary material for further discussion of the undirected edges in G3 and G4.

\begin{figure}[t]
\centering
\includegraphics[width=\linewidth]{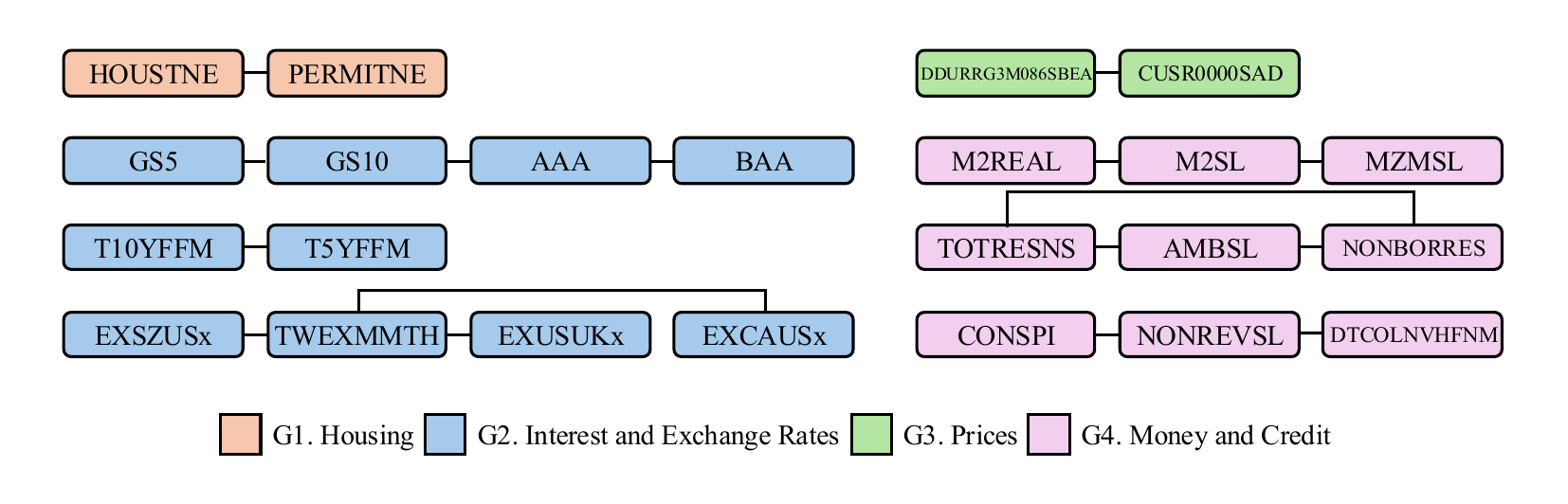}
  \caption{Estimated conditional independence graph for the FRED-MD data.}
  \label{fig:real.CIG}
\end{figure}

Figure~\ref{fig:real.ordering} presents boxplots of the estimated causal ordering across the four groups, with the detailed causal ordering
reported in Table~{\color{blue}S.2} of the supplementary material. Several well-established findings on monetary policy transmission are evident. First, the federal funds rate (FEDFUNDS) appears at the top of the estimated ordering, followed by short-term interest rates such as the 3-month commercial paper rate (CP3Mx) and the 1-year Treasury yield (GS1), and then by longer-term yields and monetary aggregates in G4. This aligns well with the interest rate channel of monetary policy transmission \citep{Bernanke1992}, which identifies the federal funds rate as the key indicator of monetary policy. Second,  the housing group (G1) lies in the middle of the ordering, which highlights its interest-sensitive nature and lends further support to the credit and balance-sheet channels \citep{Matteo2010}. Lastly, the prices group (G3) is dispersed throughout the ordering, suggesting heterogeneous price responses to monetary policy shocks \citep{ Nakamura2008}.

\begin{figure}[tbp]
\centering
\includegraphics[height=5cm]{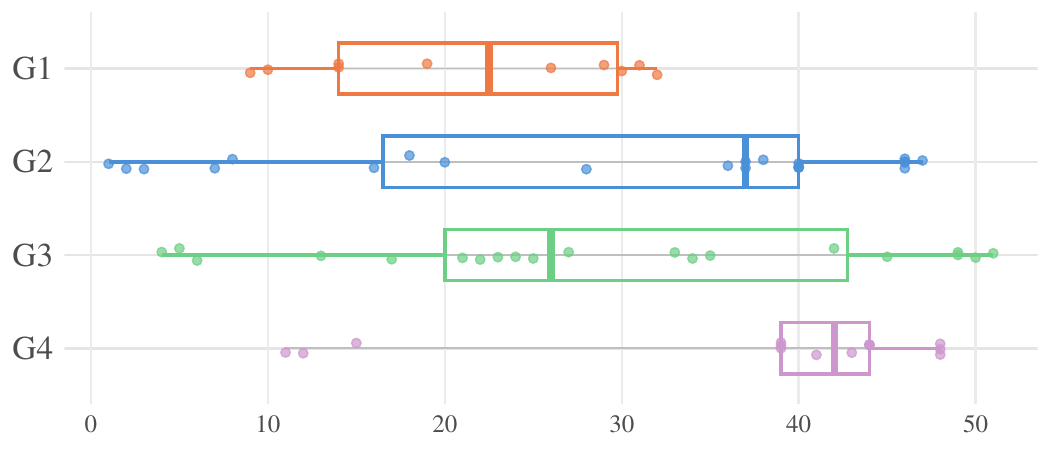}
  \caption{The boxplots of the estimated causal ordering for the FRED-MD data.}
  \label{fig:real.ordering}
\end{figure}

\begin{figure}[tbp]
  \centering
  \begin{subfigure}{0.42\linewidth}
    \centering
    \includegraphics[height=2.5cm]{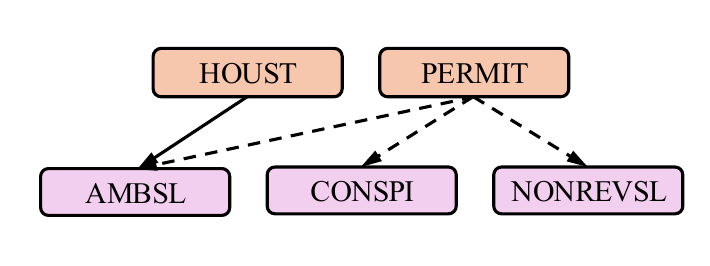}
    \vspace{-0.3cm}
    \caption{$\widehat \bA$}
  \end{subfigure}
  \hspace{1em} 
  \begin{subfigure}{0.42\linewidth}
    \centering
    \includegraphics[height=2.5cm]{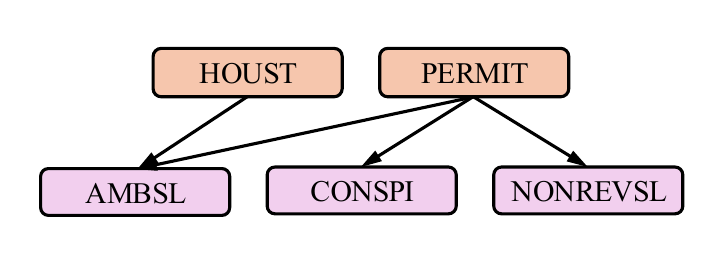}
    \vspace{-0.3cm}
    \caption{$\widehat \bB$}
  \end{subfigure}
  \caption{Common directed edges in $\widehat{\mathbf{A}}$ and $\widehat{\mathbf{B}}$, with solid lines indicating positive effects and dashed lines indicating negative effects.}
  \label{fig:real.direct.common}
\end{figure}

We further estimate the coefficient matrices, where $\widehat{\mathbf{A}}$ and $\widehat{\mathbf{B}}$ represent the contemporaneous and lagged 6-month causal relations, respectively. To facilitate visualization, we select the top 10 entries with the largest absolute values from each matrix and display the directed edges common to both $\widehat{\mathbf{A}}$ and $\widehat{\mathbf{B}}$ in Figure~\ref{fig:real.direct.common}. Specifically, we observe positive contemporaneous and lagged effects of housing starts (HOUST) on adjusted monetary base  (AMBSL).
Importantly, the housing permits series (PERMIT) exhibits negative contemporaneous effects on AMBSL, household nonrevolving credit (NONREVSL), and the nonrevolving credit-to-income ratio (CONSPI), but positive lagged effects on these same variables. This sign reversal implies that an initial increase in housing permits temporarily tightens liquidity and credit conditions, possibly due to short-term balance-sheet adjustments, but subsequently leads to an expansion as higher housing investment enhances collateral and credit growth. See also
Figure~{\color{blue}S.1} of the supplementary material for the directed edges specific to $\widehat{\mathbf{A}}$ and $\widehat{\mathbf{B}}$, and Section~{\color{blue}C} for further discussion.

To complete the analysis, we finally transform the estimated CIGs in Figure~\ref{fig:real.CIG} into Granger causality graphs.
Specifically, we apply the algorithm of \citet{songsiri2010topology} to estimate a VAR(6) model within each chain component for its residual time series, 
subject to the conditional independence constraints identified in Figure~\ref{fig:real.CIG}. See also Remarks~\ref{rmk:e_var} and \ref{rm.CIG}. For illustration, we present in Figure~\ref{fig:real.causality} the estimated Granger causality graphs for selected variables in G2. 
It is worth noting that contemporaneous conditional dependencies are found among Treasury yields and among corporate bond yields, while lagged causal effects run from Treasury yields to corporate bond yields. This pattern reveals a clear transmission of movements in government rates to corporate borrowing costs \citep{Longstaff2005}.

 \begin{figure}[tbp]
  \centering
  
   \begin{minipage}{0.45\linewidth}
   \hspace{3em}
       \begin{subfigure}{\linewidth}
      \centering
\includegraphics[height=3.3cm]{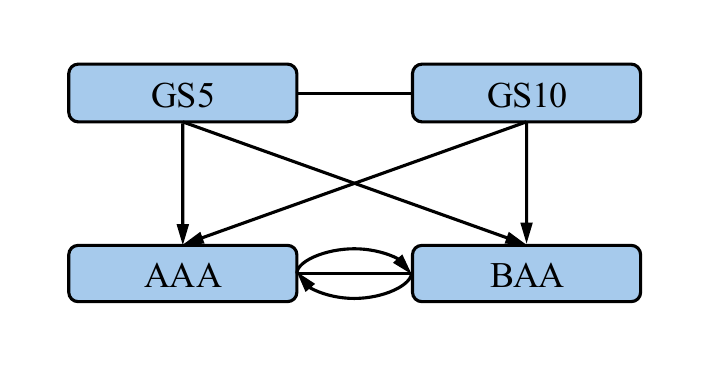}
    \end{subfigure}
    \end{minipage}
  \hfill
  \begin{minipage}{0.45\linewidth}
    \hspace{-3em}
    \begin{subfigure}{\linewidth}
      \centering
\includegraphics[height=3.3cm]{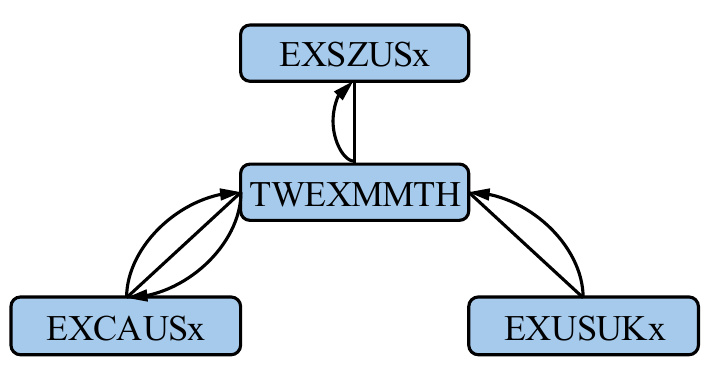}
    \end{subfigure}
  \end{minipage} 
  \caption{Estimated Granger causality graphs for selected variables in G2,  with the left and right panels respectively depicting
the interest-rate and exchange-rate chain components.}
  \label{fig:real.causality}
\end{figure}


\linespread{1.1}\selectfont
\bibliographystyle{dcu}
\bibliography{main}

\end{document}